\documentclass[aps, prd,twocolumn,superscriptaddress,nofootinbib,preprintnumbers, 10pt]{revtex4-1}
\usepackage[utf8]{inputenc}
\usepackage[english]{babel}
\usepackage{natbib}
\def\apj{ApJ}%
\def\mnras{MNRAS}%
\def\prd{Phys.~Rev.~D}%
\usepackage{graphicx}
\usepackage{amsmath,amssymb}
\usepackage{braket}
\usepackage{fontawesome}
\usepackage{float}
\usepackage[dvipsnames]{xcolor}
\usepackage{soul}
\usepackage{rotating}
\usepackage{multirow}
\usepackage{mathtools}
\usepackage{makecell}

\usepackage[caption = false]{subfig}
\usepackage{txfonts}
\usepackage[margin=0.75in]{geometry}
\usepackage[normalem]{ulem}
\usepackage{hyperref}
\hypersetup{
    colorlinks=true,
    linkcolor=red,
    citecolor=blue,
}

\DeclareMathAlphabet\mathbfcal{OMS}{cmsy}{b}{n}

\newcommand{\beq}{\begin{equation}}
\newcommand{\eeq}{\end{equation}}
\newcommand{\beqa}{\begin{eqnarray}}
\newcommand{\eeqa}{\end{eqnarray}}

\definecolor{darkgreen}{rgb}{0.0, 0.5, 0.0}
\definecolor{darkcyanxf}{RGB}{0.0, 139.0, 139.0}

\newcommand{\planck}{\textit{Planck}}
\newcommand{\act}{\textsc{ACT}}

\begin{document}
\title[Shear-y]{Constraints on cosmology and baryonic feedback with joint analysis of Dark Energy Survey Year 3 lensing data and ACT DR6 thermal Sunyaev-Zel'dovich effect observations}

\label{firstpage}

\begin{abstract}
We present a joint analysis of weak gravitational lensing (shear) data obtained from the first three years of observations by the Dark Energy Survey and thermal Sunyaev-Zel'dovich (tSZ) effect measurements from a combination of Atacama Cosmology Telescope (ACT) and \planck\ data. A combined analysis of shear (which traces the projected mass) with the tSZ effect (which traces the projected gas pressure) can jointly probe both the distribution of matter and the thermodynamic state of the gas, accounting for the correlated effects of baryonic feedback on both observables. We detect the shear$~\times~$tSZ cross-correlation at a 21$\sigma$ significance, the highest to date, after minimizing the bias from cosmic infrared background leakage in the tSZ map. By jointly modeling the small-scale shear auto-correlation and the shear$~\times~$tSZ cross-correlation, we obtain $S_8 = 0.811^{+0.015}_{-0.012}$ and $\Omega_{\rm m} = 0.263^{+0.023}_{-0.030}$, results consistent with primary CMB analyses from \planck\ and P-ACT. We find evidence for reduced thermal gas pressure in dark matter halos with masses $M < 10^{14} \, M_{\odot}/h$, supporting predictions of enhanced feedback from active galactic nuclei on gas thermodynamics. A comparison of the inferred matter power suppression reveals a $2-4\sigma$ tension with hydrodynamical simulations that implement mild baryonic feedback, as our constraints prefer a stronger suppression. Finally, we investigate biases from cosmic infrared background leakage in the tSZ-shear cross-correlation measurements, employing mitigation techniques to ensure a robust inference. Our code is publicly available on GitHub \href{https://github.com/shivampcosmo/GODMAX/tree/DESxACT}{\textcolor{black}{\faGithubSquare}}.\footnote{\url{https://github.com/shivampcosmo/GODMAX/tree/DESxACT}}   
\end{abstract}

\author{S.~Pandey}\email{shivampcosmo@gmail.com}
\affiliation{William H.~Miller III Department of Physics \& Astronomy, Johns Hopkins University, Baltimore, MD 21218, USA}
\affiliation{Department of Physics, Columbia University, New York, NY, USA 10027}

\author{J.~C.~Hill}
\affiliation{Department of Physics, Columbia University, New York, NY, USA 10027}

\author{A.~Alarcon}
\affiliation{Institute of Space Sciences (ICE, CSIC), Campus UAB, Carrer de Can Magrans, s/n, 08193 Barcelona, Spain}
\affiliation{Argonne National Laboratory, 9700 South Cass Avenue, Lemont, IL 60439, USA}
\author{O.~Alves}
\affiliation{Department of Physics, University of Michigan, Ann Arbor, MI 48109, USA}
\affiliation{Instituto de F\'{i}sica Te\'orica, Universidade Estadual Paulista, S\~ao Paulo, Brazil}
\affiliation{Laborat\'orio Interinstitucional de e-Astronomia, Rua Gal. Jos\'e Cristino 77, Rio de Janeiro, RJ - 20921-400, Brazil}
\author{A.~Amon}
\affiliation{Department of Astrophysical Sciences, Princeton University, Peyton Hall, Princeton, NJ 08544, United States of America}

\author{D.~Anbajagane}
\affiliation{Department of Astronomy and Astrophysics, University of Chicago, Chicago, IL 60637, USA}

\author{F.~Andrade-Oliveira}
\affiliation{Instituto de F\'{i}sica Te\'orica, Universidade Estadual Paulista, S\~ao Paulo, Brazil}
\affiliation{Laborat\'orio Interinstitucional de e-Astronomia, Rua Gal. Jos\'e Cristino 77, Rio de Janeiro, RJ - 20921-400, Brazil}
\author{N.~Battaglia}
\affiliation{Department of Astronomy, Cornell University, Ithaca, NY 14853, USA}
\affiliation{Université Paris Cité, CNRS, Astroparticule et Cosmologie, F-75013 Paris, France}
\author{E.~Baxter}
\affiliation{Department of Physics and Astronomy, Watanabe 416, 2505 Correa Road, Honolulu, HI 96822}
\author{K.~Bechtol}
\affiliation{Physics Department, 2320 Chamberlin Hall, University of Wisconsin-Madison, 1150 University Avenue Madison, WI  53706-1390}
% ...

\maketitle

\section{Introduction}\label{sec:intro}
As baryons interact via forces other than gravitation, they cool to form compact objects like stars and black holes. When stars explode as supernovae, or black holes turn into active galactic nuclei, they can output large amounts of energy that can impact large-scale structure (LSS) even at several megaparsec (Mpc) scales. The gas ejected out of a dark matter halo feeds onto the large-scale structure through cosmic web, creating a coupled feedback loop by getting accreted onto neighboring halos and impacting their behavior during the evolution of the Universe. However, due to the large dynamical range of scales involved, these baryonic feedback processes are difficult to understand from first principles, making it one of the leading sources of theoretical uncertainty in modeling the LSS (see, e.g., \cite{chisari_modelling_2019} and references therein). 

Cross-correlating cosmic probes that are impacted by baryonic feedback in different ways presents an opportunity to understand baryonic processes using a data-driven approach. In this study, we focus on the weak gravitational lensing of background galaxies (see \citep{kilbinger_cosmology_2015} for a review) and the inverse-Compton scattering of the cosmic microwave background (CMB) by hot electrons in the LSS (see \citep{carlstrom_cosmology_2002, birkinshaw_sunyaev-zeldovich_1999} for a review). Weak lensing, also called cosmic shear, measures the correlated deformation of shapes of galaxies and is sensitive to the integrated distribution of the total matter (i.e. dark and baryonic). In contrast, the inverse-Compton scattering of the CMB off ionized electrons moving with random thermal motions, also called the thermal Sunyaev-Zel'dovich (tSZ) effect \citep{sunyaev_small-scale_1970}, is sensitive to the integrated pressure along a line of sight of the hot gas in the Universe. A related effect, caused by the bulk flow of gas, causes the kinetic Sunyaev-Zel'dovich (kSZ) effect and is sensitive to the integrated electron momentum along the same line of sight. Joint modeling of such cross-correlations can help constrain the physics of baryonic feedback and how it impacts the distribution and thermodynamics of the baryons as well as its back-reaction on the dark matter distribution. 

In recent years, the tSZ signal has been cross-correlated with both galaxies \citep{kou_testing_2024, liu_measurements_2025, tramonte_exploring_2023, pandey_constraints_2019, koukoufilippas_tomographic_2020, makiya_joint_2018, chiang_cosmic_2020, schaan_atacama_2021, lokken_superclustering_2025} and weak lensing fields \citep{hill_detection_2014, hojjati_cross-correlating_2017, osato_cross-correlation_2020, gatti_cross-correlation_2022, troster_joint_2022, mccarthyhill2024}. The large-scale analyses of such cross-correlations have placed powerful constraints on the average (bias weighted-) pressure of the hot gas (Eq.~\ref{eq:bA}) in the Universe and its evolution with redshift. With the higher-resolution microwave observatories such as the Atacama Cosmology Telescope (ACT) \cite{henderson_advanced_2016} and the South Pole Telescope (SPT) \citep{bleem_galaxy_2015}, it has become possible to explore small-scale correlations, probing gas properties inside the halos \citep{amodeo_atacama_2021, sanchez_mapping_2023, pandey_cross-correlation_2022, troster_joint_2022, liu_measurements_2025}. 

While tSZ, lensing, and galaxies all trace the same underlying LSS, most studies model these probes with independent parameterizations. 
In \cite{hojjati_cross-correlating_2017}, the authors assumed the universal pressure profile \citep{arnaud_universal_2010} for gas pressure and the Navarro-Frenk-White (NFW, \citep{navarro_universal_1997}) profile for dark matter distribution to analyze the shear $~\times~$ tSZ cross-correlations. However, due to baryonic feedback, we expect the expelled gas to alter the total matter profile from the NFW fitting function. In \cite{gatti_cross-correlation_2022, pandey_cross-correlation_2022}, the authors measured and analyzed the shear$~\times~$tSZ cross-correlations from DES, ACT, and \textit{Planck} datasets to additionally allow for bloating of the halo profiles and changes to the halo concentration due to the evolution of baryonic processes \citep{mead_accurate_2015}. However, we physically expect tSZ and lensing probes to be highly correlated in their properties and evolution.

In \cite{mead_hydrodynamical_2020}, the authors developed a connected hydrodynamical model, which was used to analyze the weak lensing and tSZ data from KiDS and \planck\ in \cite{troster_joint_2022}. However, this model makes simple approximations in relating the gas density and pressure profiles, such as imposing a polytropic form for the gas density, which can be violated in lower-mass halos \citep{capelo_polytropic_2012, battaglia_cluster_2012}. In \cite{pandey_accurate_2025}, the authors developed a more flexible model, relating baryonic thermodynamics and the matter distribution, solving exact hydrostatic equilibrium equations that relate the gas density and pressure as well as account for realistic non-thermal pressure support in halos. This model was then validated on a large suite of hydrodynamical simulations with highly varying feedback strengths (over halos in mass and redshift ranges of $10^{13} < {\rm M} \, (M_{\odot}/h) < 10^{15}$ and $0 < z < 1$), to show that the model is flexible enough to jointly fit the gas density, gas pressure, and total matter density profiles in a wide range of halo masses and redshifts. 

In this study, we use the model from \cite{pandey_accurate_2025} to jointly analyze the tSZ and weak lensing fields. In contrast, previous analyses by \cite{gatti_cross-correlation_2022, pandey_cross-correlation_2022} combined high-resolution ACT DR4 maps on a limited 400 $\rm deg^2$ field (the \texttt{D56} region) with the lower-resolution \planck\ tSZ map (with a $\sim~6$ times larger beam compared to ACT) for the remaining DES area. Here, we use the full ACT DR6 dataset, which provides uniform, high-resolution coverage across the entire $\sim$4000 $\rm deg^2$ DES footprint. This provides sensitivity to lower-mass halos and smaller-scale gas profiles, which is crucial for constraining AGN feedback activity.

The shear$~\times~$tSZ cross-correlations are complementary to galaxy$~\times~$tSZ and tSZ$~\times~$tSZ correlations as well as to other thermodynamic probes, such as X-ray and kSZ. The thermodynamic properties of group-scale halos ($M \lesssim 5 \times 10^{13} M_{\odot}/h$) can be probed with galaxy-tSZ and galaxy-kSZ cross-correlations, whereas the tSZ auto-power spectra and X-ray observations probe cluster-scale halos ($M \gtrsim 5\times 10^{14} M_{\odot}/h$). However, the shear$~\times~$tSZ cross-correlation is sensitive to intermediate-mass halos, $5 \times 10^{13} \lesssim M (M_{\odot}/h) \lesssim 5 \times 10^{14}$ \citep{battaglia_deconstructing_2015, hojjati_dissecting_2015, osato_cross-correlation_2020, pandey_cross-correlation_2022}. Constraining the gas thermodynamics in this intermediate mass range is lucrative because AGNs (residing in these halos) are efficient in pushing the gas out of the halos but with relatively less stochasticity compared to low-mass halos. Note that the shear 2-point auto-correlation is also sensitive to the total matter distribution in halos in a similar mass range \citep{to_deciphering_2024}. Therefore, a data-driven constraint on the gas profile in these halos obtained using SZ effect can help in calibrating the baryonic feedback effects in the shear 2-point auto-correlation and improve cosmological constraints \citep{van_daalen_effects_2011, chisari_modelling_2019}.

The sky observations at microwave wavelengths made by the \planck\ and \act\ observatories include signals from not only the CMB and the tSZ effect, but also relatively poorly understood emissions from dusty galaxies that produce the cosmic infrared background (CIB). A common approach to extracting a map of the desired signal (e.g., tSZ or the blackbody CMB) is component separation, a technique that uses multi-frequency observations of the microwave sky to build a map that ensures unit response to the spectral energy distribution (SED) of the component of interest and minimizes noise; the method can be extended to deproject a contaminant(s) with a given SED \citep{delabrouille_full_2009,remazeilles_cmb_2011}. Given our limited understanding of the properties and spectral coverage of dusty galaxies that source the CIB, robustly removing this signature from the tSZ maps remains challenging \citep{surrao2025}. We discuss its impact on the shear$~\times~$tSZ cross-correlations below. 

The remainder of this paper is organized as follows. In \S~\ref{sec:meas_mod} we describe the data products and measurement methodology. We also describe the theory model and analysis settings used to interpret these measurements. In \S~\ref{sec:results} we present the cosmological and astrophysical constraints obtained from our analysis. In \S~\ref{sec:conclusion}, we discuss the implications of our findings and conclude.

\begin{figure*}
    \centering
    \subfloat[]{
        \includegraphics[width=0.445\linewidth]{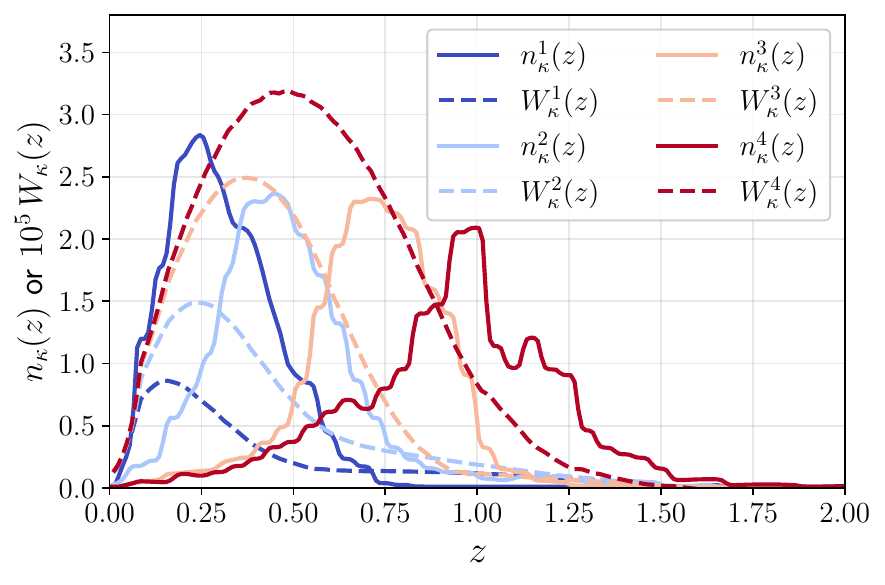}
    }
    \hfill
    \subfloat[]{%
        \includegraphics[width=0.544\linewidth]{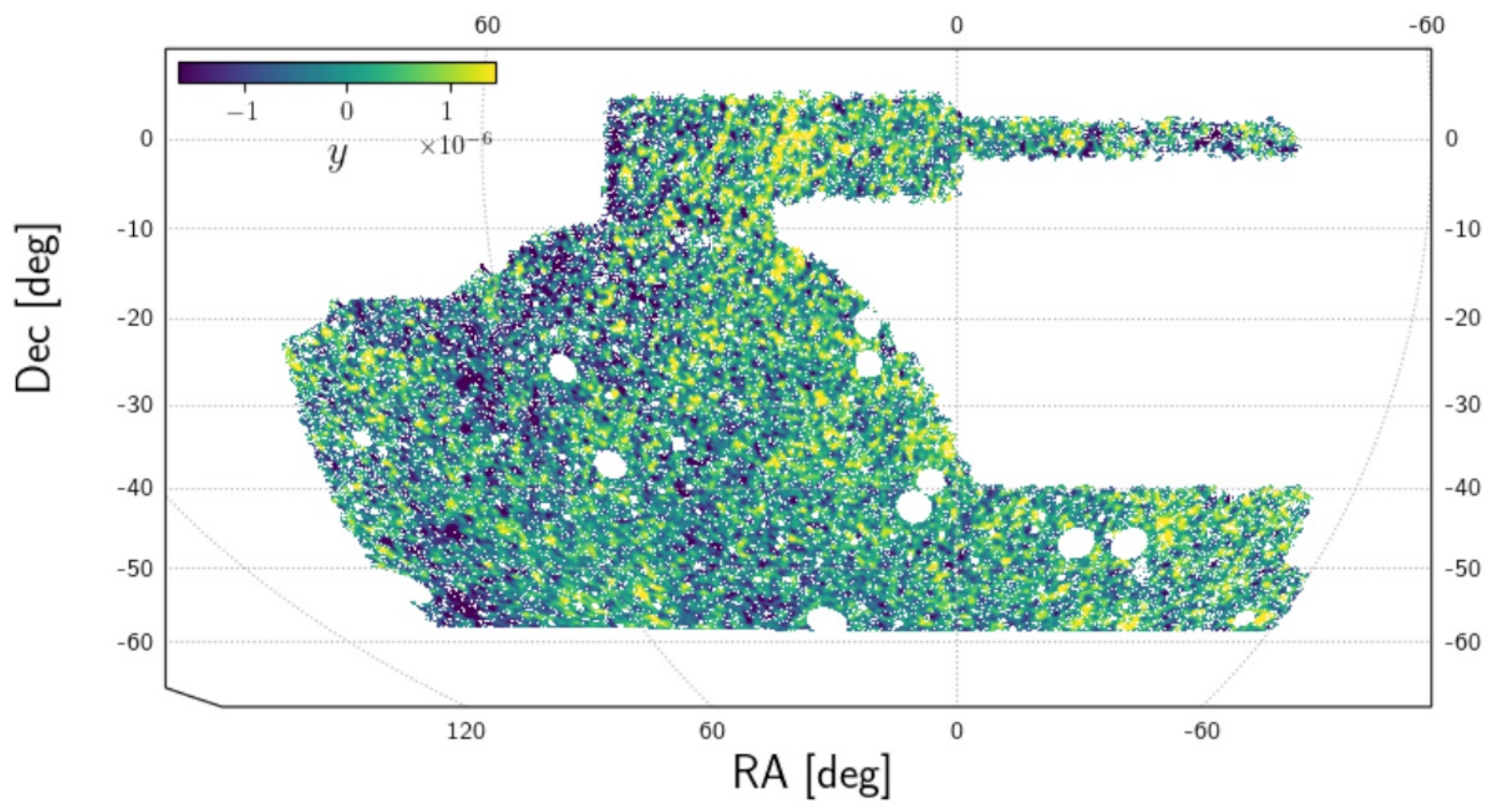}
    }
    \caption{Left panel: the redshift distributions of the shear catalog from DES split into four tomographic bins shown with solid lines. We also show the corresponding lensing efficiency (with dashed curves) for each tomographic bin (see Eq.~\ref{eq:Wk}). 
    Right panel: the fiducial tSZ map (CIB SED moment deprojected) from ACT + \emph{Planck} on the common ACT $\times$ DES footprint.}
    \label{fig:Wk_nz_tSZ_map}
\end{figure*}

\begin{figure*}
\centering
\includegraphics[width=\textwidth]{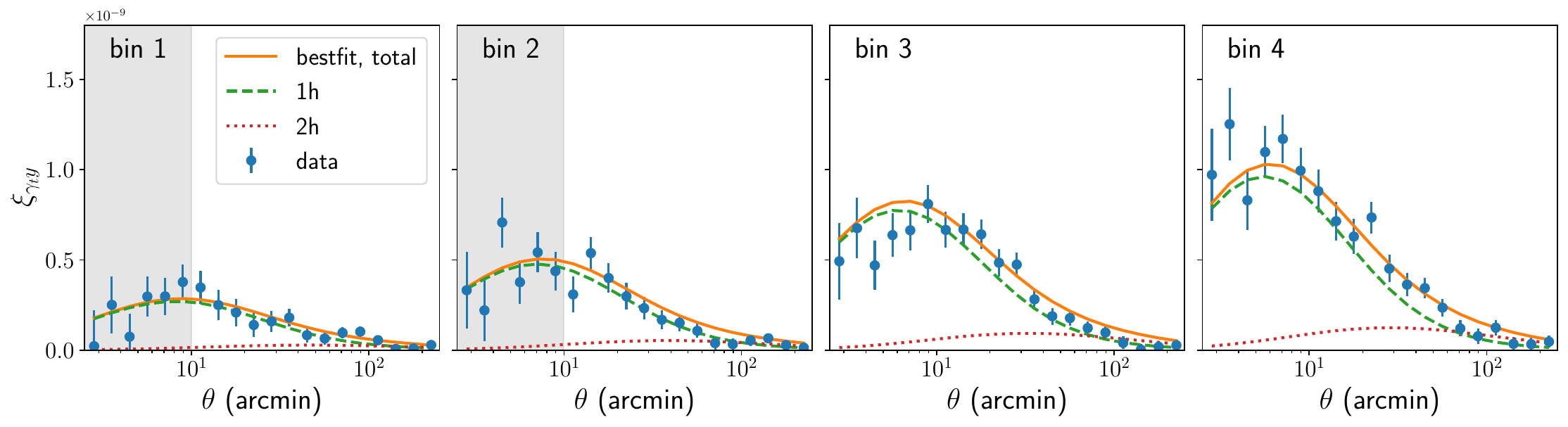}
\caption[]{Measurements of the shear-tSZ cross-correlation, $\xi_{\gamma_t y}$, using the four tomographic bins of the DES Y3 shear catalog and the CIB moment-deprojected Compton-$y$ map from ACT. Each tomographic bin probes a different redshift range, as shown by the lensing efficiency curves in the left panel of Fig.~\ref{fig:Wk_nz_tSZ_map}. The solid lines show the best-fit model, while the dashed and dotted lines show its decomposition into the 1-halo and 2-halo contributions, respectively (see \S~\ref{sec:Pk_xi}). The shaded regions denote scales excluded from this analysis because they are affected by higher-order intrinsic alignment terms not included in our fiducial model (see \S~\ref{sec:IA}).
}
\label{fig:xigty_measure}
\end{figure*}

\section{Measurements, modeling, and analysis}
\label{sec:meas_mod}

\subsection{Measurements}
\label{sec:measurement}

We use the Dark Energy Survey (DES) Year 3 (Y3) shape catalog, as described in \cite{gatti_dark_2021}, which tabulates the shapes of more than 100 million galaxies over an area of more than 4100 square degrees, giving an effective density of 5.6 galaxies/${\rm arcmin}^2$. The shapes of the galaxies are inferred using the \texttt{metacalibration} pipeline \citep{huff_metacalibration_2017, sheldon_practical_2017}, which accounts for various observational and astrometric effects \cite{sevilla-noarbe_dark_2021, jarvis_dark_2021}. However, the impacts of shear-dependent biases and object blending are calibrated using image simulations \citep{maccrann_dark_2022}, characterized by shear calibration parameters. These galaxies are divided into four tomographic bins using the methodology described in \cite{myles_dark_2021}. The redshift distribution for these four bins is shown in the left panel of Fig.~\ref{fig:Wk_nz_tSZ_map}. We marginalize over uncertainties in the mean of the redshift distribution for each tomographic bin ($\Delta z^{i}$), as well as shear calibration ($m^i$), using Gaussian priors (see Appendix~\ref{app:all_params}), similar to the fiducial DES Y3 cosmology analysis \cite{des_collaboration_dark_2022}.

The tSZ map used in this analysis, obtained from combined observations of ACT and \planck, is described in \cite{coulton_atacama_2024}. Specifically, we use data from ACT DR4 and DR6 \citep{naess_atacama_2025, aiola_atacama_2020}, which include observations in three frequency bands centered at approximately 90, 150, and 220 GHz, along with \planck\ data~\cite{akrami_planck_2020}. The single-frequency maps from both ACT and \planck\ in the common DR6 footprint are combined using a component separation pipeline based on needlet decomposition \cite{delabrouille_full_2009} to robustly extract the tSZ signal, while deprojecting various contaminants as described below. Note that the individual frequency maps are convolved with a common beam of full width at half maximum (FWHM) 1.6~arcmin. We additionally mask out the regions where individual bright point sources (such as radio galaxies) were detected above a threshold of roughly 4-10 mJy \citep{naess_atacama_2025, louis_atacama_2025} (see Appendix~\ref{app:cib} for details). 
The tSZ map used in this analysis, covering the $f_{\rm sky} = 0.087$ common footprint between DES and ACT, is shown in the right panel of Fig.~\ref{fig:Wk_nz_tSZ_map}. As described in \S~\ref{sec:intro}, leakage from the CIB into a component-separated tSZ map can bias derived cross-correlation statistics, such as shear$~\times~$tSZ. We therefore use the CIB-cleaned maps from \cite{coulton_atacama_2024}, which are provided in the \texttt{CAR} format and were produced using a moment deprojection method detailed below and in Appendix~\ref{app:cib}.

We follow the methodology described in \cite{gatti_cross-correlation_2022} to measure the shear$~\times~$tSZ cross-correlation, $\xi_{\gamma_t y}$. The estimator is approximately given by $\xi_{\gamma_t y} = \sum_{\rm pixels} y \, w_{S} e_{t}/\sum w_{S}$, where $w_{S}$ are the systematic weights, $e_{t}$ is the tangential ellipticity of source galaxies around a central pixel, and $y$ is the tSZ value of the pixel. We use \texttt{treecorr} \cite{jarvis_treecorr_2015} to measure the cross-correlation in 20 angular bins logarithmically spaced between 2.5 and 250 arcmin, which is the same binning as used by the DES Y3 analysis \cite{des_collaboration_dark_2022}. In Fig.~\ref{fig:xigty_measure} we show the measurements of the shear-\textit{y} correlations. Additionally, we show the best-fit curve, as well as its decomposition into 1-halo and 2-halo components, using the model described in the following subsection. In Appendix~\ref{app:shear2pt}, we similarly show the measurement and best-fit curves for the shear two-point correlation. Note that the measurements there are obtained with the same method as presented in \cite{amon_dark_2022, secco_dark_2022}\footnote{Note that there was an update to the redshift assignment of the source galaxies where the binning was updated to v0.5 of SOMPZ (see \url{https://des.ncsa.illinois.edu/releases/y3a2/Y3key-catalogs}). We use this updated binning in this work which changes the measurements slightly. See \citep{To_obs:25} for more details. }.

As mentioned earlier, the CIB contaminates the tSZ map constructed through component separation. Since the CIB is sourced by dusty galaxies, which trace the large-scale structure, it can bias the shear$~\times~$tSZ cross-correlations. In \cite{coulton_atacama_2024}, the authors generate an array of maps under different assumptions about the SED of dusty galaxies. However, it was shown that if, in addition to deprojecting the CIB-modified blackbody spectral index ($\beta$), we also deproject its first moment (henceforth referred to as $\beta + d\beta$), it results in a robust tSZ map that remains insensitive to assumptions about the value of $\beta$ \citep{chluba_rethinking_2017, coulton_atacama_2024}. We confirm these findings, as described in Appendix~\ref{app:cib}. Since the number of observed frequency bands is limited, deprojecting additional components increases noise. We find that without any deprojection, the shear$~\times~$tSZ cross-correlation is measured at 25$\sigma$ significance, which drops to approximately $21\sigma$ with moment deprojection. However, since the $\beta + d\beta$ moment-deprojected maps are more robust, we choose them for our fiducial results (obtained at $\beta=1.7$).

\subsection{Modeling}\label{sec:model}

We use the framework described in \cite{pandey_accurate_2025} to model the shear$~\times~$tSZ and shear auto-correlations in this work. This approach uses an analytical halo model and consistently models the correlated spherical profiles of various components by generalizing the prescription described in \cite{schneider_quantifying_2019} and adding consistent predictions for the baryon thermodynamics, including thermal and non-thermal pressure. For computational feasibility, we assume that the halo concentration follows the mean concentration-mass relation of \cite{duffy_dark_2008}. Since both the shear$~\times~$tSZ and shear auto-correlations are sensitive to a broad range of halo masses and redshifts, we find this to be a good approximation. Other assumptions and caveats to the model are described in the following text. This model has been implemented using the \texttt{JAX}\footnote{\url{https://jax.readthedocs.io}} library which offers automatic differentiation of various \texttt{Python} functions, out-of-the-box parallelization schemes, and just-in-time compilation for both CPU and GPU. The automatic differentiation functionality enables interfacing with efficient sampling schemes like Hamiltonian Monte Carlo that we use to explore the parameter space \citep{duane_hybrid_1987, neal_mcmc_2011}.

\subsubsection{Profiles}
The density profile of a halo in a dark matter-only Universe is given by the Navarro-Frenk-White (NFW) profile ($\rho_{\rm nfw}$) \citep{navarro_universal_1997}.\footnote{Following \cite{oguri_detailed_2011}, the NFW profile, $\rho_{\rm nfw}$, is truncated at the halo outskirts using a damping term.} The normalization of this profile is fixed by ensuring that the total mass of the halo within radius $r_{\rm 200c}$ equals $M_{\rm 200c}$.\footnote{Note that the spherical overdensity radius, $r_{200c}$, of a halo at redshift $z$ is defined such that the average enclosed density within a sphere of radius $r_{\rm 200c}$ is equal to $200$ times the critical density of the Universe, $\rho_c(z)$: $M_{\rm 200c} = (4\pi/3) \, 200 \, r_{\rm 200c}^3 \, \rho_c(z)$, where the spherical overdensity mass, $M_{\rm 200c}$, is the mass within $r_{\rm 200c}$.}
However, in a Universe with both dark matter and baryons, the total matter density profile of this halo can be split into three components: $\rho_{\rm dmb}(r) = \rho_{\rm cga}(r) + \rho_{\rm gas}(r) + \rho_{\rm clm}(r)$, where $\rho_{\rm cga}(r)$ is the stellar mass profile of the central galaxy, $\rho_{\rm gas}(r)$ is the gas mass profile, and $\rho_{\rm clm}(r)$ is the profile of collisionless matter (including the dark matter and the stellar mass of satellite galaxies). We refer the readers to \cite{pandey_accurate_2025} for details on the implementation of the stellar profiles used in this paper. We assume a conserved total halo mass, $M_{\rm tot}$, given by $M_{\rm tot} = 4\pi \int_0^{\infty} dr \, r^2 \, \rho_{\rm dmb}(r) = 4\pi \int_0^{\infty} dr \, r^2 \, \rho_{\rm nfw}(r)$.\footnote{While $M_{\rm tot}$ is conserved by construction, other mass definitions like $M_{\rm 200c}$ can change due to gas expulsion from the halo via baryonic feedback.}  

The gas density profile is parameterized as \citep{giri_emulation_2021, pandey_accurate_2025}:
\begin{equation}\label{eq:rho_gas}
\rho_\mathrm{gas}(r) = \frac{\rho_{\rm gas, 0}}{\left[1+ \left(\frac{r}{\theta_{\rm co} r_\mathrm{200c}}\right) \right]^{\beta_{\rm gas}} \left[1+ \left(\frac{r}{\theta_\mathrm{ej}r_\mathrm{200c}}\right)^{\gamma_{\rm gas}} \right]^{\frac{\delta_{\rm gas}-\beta_{\rm gas}}{\gamma_{\rm gas}}}},
\end{equation}
where the parameters $\theta_{\rm co}$ and $\theta_{\rm ej}$ control the core and ejection radii of the gas, respectively. The parameters $\beta_{\rm gas}$, $\gamma_{\rm gas}$, and $\delta_{\rm gas}$ determine the slope of the profile. These parameters account for the impact of baryonic feedback, which typically ejects gas from the halo’s interior to its outskirts, leading to a profile that can deviate significantly from the NFW profile by developing a central core and a shallower slope in the outskirts. As the parameter space is substantially degenerate, we fix $\gamma_{\rm gas} = 2.0$, $\delta_{\rm gas} = 7.0$, and $\theta_{\rm co} = 0.05$ in this analysis, while varying the parameters controlling $\beta_{\rm gas}$ and $\theta_{\rm ej}$. These default values of fixed parameters and sampled parameter space was found sufficient to jointly fit the matter density and pressure profiles of halos in mass and redshift ranges of $13.0 < \log(M) < 14.5$ and $0 < z < 1$ in the hydrosimulations in the ANTILLES simulation suite \citep{salcido_spk_2023} with varying the baryonic feedback strength \citep{pandey_accurate_2025}. The normalization factor $\rho_{\rm gas, 0}$ is determined by enforcing that the total gas mass out to large radii is $M_{\rm gas}(< \infty) = f_{\rm gas} M_{\rm tot}$, where $f_{\rm gas} = \Omega_{\rm b}/\Omega_{\rm m} - f_{\rm star}$ represents the universal gas fraction.

As low-mass halos have shallower gravitational potential wells, baryonic feedback from AGNs (which evolves with redshift) becomes more efficient in pushing the gas out of the halo \citep{wright_baryon_2024}. Therefore, we expect the spatial extent of ejected gas to evolve with halo mass and redshifts which, following \cite{pandey_accurate_2025}, we parametrize as follows:
\begin{equation}\label{eq:theta_ej_co}
    \theta_{\rm ej} = \theta_{\rm ej,0} \bigg(\frac{M_{\rm 200c}}{M_{\rm \star, ej}}\bigg)^{\nu^{M}_{\theta_{\rm ej}}} \, (1 + z)^{\nu^z_{\rm ej}},
\end{equation}
where we treat $\theta_{\rm ej,0}$, $\nu^{M}_{\theta_{\rm ej}}$, and $\nu^z_{\rm ej}$ as free parameters, while setting $\log_{10}(M_{\rm \star, ej}) = 16$. Note that we use a broad uniform prior on all of the three free parameters (see Table~\ref{tab:params_all}). Our priors enforce the condition that for the most massive halos ($\log_{10}(M_{200c}) \sim 16$), gas is ejected to less than $6\, R_{200c}$, which is expected even in high baryonic feedback simulations due to their deep gravitational potential wells \citep{schaye_flamingo_2023}. 

We parameterize the evolution of $\beta$ with mass similar to \citep{giri_emulation_2021}:
\begin{equation}\label{eq:beta_gas}
    \beta_{\rm gas} = \frac{3(M_{\rm 200c}/M_{\rm c})^{\mu_{\beta}}}{1+(M_{\rm 200c}/M_{\rm c})^{\mu_{\beta}}} \ ,
\end{equation}
which allows for the gas profile to become shallower than the NFW profile in low-mass halos ($M_{\rm 200c} < M_{c}$), as expected from AGN feedback. In this form, we also physically expect $\mu_{\beta} > 0$ \citep{giri_emulation_2021} (Table~\ref{tab:params_all}) and we fix $\log_{10}(M_c) = 13.75$.

Given the gas density and stellar profiles, we solve for $\rho_{\rm clm}(r)$ by approximately conserving angular momentum, as detailed in \cite{schneider_quantifying_2019, abadi_galaxy-induced_2010}.

Using the total matter density profile, we solve for the total pressure profile, $P_{\rm tot}(r)$, by applying the hydrostatic equilibrium equation. The fraction of non-thermal pressure support in the total pressure is given by \citep{shaw_impact_2010, osato_baryon_2023}:
\begin{equation}\label{eq:Pnt}
  R_\mathrm{nt} = \frac{P_\mathrm{nt}}{P_\mathrm{tot}} = \alpha_\mathrm{nt} f(z)
  \left( \frac{r}{r_{\rm 200c}} \right)^{n_\mathrm{nt}},
\end{equation}
where $\alpha_{\rm nt}$ sets the amplitude of non-thermal pressure support, $f(z)$ governs its redshift evolution, and $n_{\rm nt}$ sets its radial dependence. We set $n_{\rm nt} = 0.3$ and $f(z)$ as prescribed by \cite{shaw_impact_2010}, while varying $\alpha_\mathrm{nt}$ with a broad prior (see Table~\ref{tab:params_all}). Some hydrodynamical simulations \citep{Battaglia:2012:ApJ:1} find that the non-thermal pressure fraction has a mild dependence on halo mass and redshift, and that its radial profile deviates from the simple power-law model considered here. We defer a detailed implementation and analysis of these additional degrees of freedom to future work. With this phenomenological model for the non-thermal pressure support, we compute the thermal pressure profile as $P_\mathrm{th} = P_\mathrm{tot} \times \mathrm{max} \left[ 0, 1 - R_\mathrm{nt} \right]$. The electron pressure profile that generates the thermal SZ signal is given by $P_e(r) = \frac{2(X_{\rm H} + 1)}{(5X_{\rm H} + 3)} P_{\rm th}$, where $X_{\rm H} = 0.76$ is the primordial hydrogen mass fraction.

\subsubsection{Power spectrum and correlation functions}\label{sec:Pk_xi}
Using the halo model framework (see \cite{cooray_halo_2002} for a review), we can express the matter power spectrum as a sum of the 1-halo and 2-halo terms. The 1-halo term can be written as:
\begin{align}
    P_{AB}^{\rm 1h}(k,z) = \int_{M_{\rm min}}^{M_{\rm max}} dM \frac{dn}{dM}  u_A (k,z,M) \, u_B (k,z,M),
\end{align}
where $u_A$ and $u_B$ are the Fourier-space profiles of the probes $A$ and $B$, and $dn/dM$ is the halo mass function, for which we use the fitting function described in \cite{tinker_large-scale_2010}. 

The 2-halo term is given by:
\begin{equation}
    P_{AB}^{\rm 2h}(k,z) = b_{A}(k, z) \, b_{B}(k, z) \, P_{\rm lin}(k,z),
\end{equation}
where $b_{A}$ and $b_{B}$ are the scale-dependent bias of the probes $A$ and $B$ respectively:
\begin{equation}\label{eq:bA}
    b_{A}(k,z) = \int_{M_{\rm min}}^{M_{\rm max}} dM \frac{dn}{dM} \,  u_A (k,z,M) \, b_{\rm halo}(M, z),
\end{equation}
with $b_{\rm halo}(M, z)$ being the large scale halo bias for which we use the fitting formulae described in \cite{tinker_large-scale_2010} and, $P_{\rm lin}(k,z)$ is the linear matter power spectrum for any given cosmology. To maintain compatibility with \texttt{JAX}, we use the formulation described in \cite{eisenstein_baryonic_1998} (as implemented in \cite{campagne_jax-cosmo_2023}) to obtain the linear power spectrum using analytical approximation. As we probe the projected fields, this approximation holds at better than 2\% level for the scales of our interest. 
However, in the future, this can be refined by interfacing with other emulators, such as \cite{bartlett_precise_2024, sui_syren-new_2024, hahn_disco-dj_2024, piras_future_2024}. Note that, for matter, we enforce mass conservation by ensuring that the consistency relation $\int b_{\rm halo} (M, z) \frac{dn}{dM} dM = 1$ holds \cite{cacciato_combining_2012, mead_hydrodynamical_2020, bolliet_projected-field_2023}. 

We can then write $P_{AB}^{\rm{tot}} = P_{AB}^{\rm 1h} + P_{AB}^{\rm 2h}$. However, this approach suffers from issues in the 1-halo to 2-halo transition regime, as it is difficult to model the halo exclusion effects analytically. To solve this, we use the response approach \cite{mead_hydrodynamical_2020, cacciato_combining_2012}. Within this ansatz, we estimate the matter power spectrum $P_{\rm mm}^{\rm halofit}(k,z)$ for the same cosmological parameters as used for $P_{AB}^{\rm{tot}}$, obtained by fitting dark-matter-only N-body simulations \citep{takahashi_revising_2012}. We also compute $P_{\rm mm}^{\rm{tot, nfw}}$ by setting the total matter distribution to be the same as the NFW profile \citep{navarro_universal_1997}, $\rho_{\rm dmb} = \rho_{\rm NFW}$. As $P_{\rm mm}^{\rm halofit}(k,z)$ is obtained by directly fitting the total power spectra in N-body simulations, it does not suffer from exclusion effects, whereas $P_{\rm mm}^{\rm{tot, nfw}}$ does.
Therefore, we compute the ratio $R(k,z) = P_{\rm mm}^{\rm halofit}(k,z)/P_{\rm mm}^{\rm{tot, nfw}}(k,z)$ and multiply it by $P_{\rm mm}^{\rm{tot}}$ to obtain the final estimate of the 3D matter power spectrum, which accounts for halo model misspecification in the transition regime. As the pressure profile differs significantly compared to the total matter profile, it has different exclusion effects compared to matter and hence we account for the transition regime misspecification by smoothing the 1-halo and 2-halo terms in the power spectrum \cite{mead_accurate_2015}, $P_{\rm ym}^{\rm tot} = \left( \left( P_{\rm ym}^{\rm 1h} \right)^{\alpha_{\kappa y}} + \left( P_{\rm ym}^{\rm 2h}{} \right)^{\alpha_{\kappa y}} \right)^{1/\alpha_{\kappa y}}$, where $\alpha_{\kappa y}$ is treated as a free parameter with uninformed prior (Table~\ref{tab:params_all}).

The 2D multipole power spectrum, $C^{ij}_{AB}(\ell)$, between probes can be obtained with a Limber integration \citep{limber_analysis_1953}:
\begin{align}
    C^{ij}_{AB}(\ell) = \int d
    \chi \,P_{AB}^{\rm{tot}}\bigg( k=\frac{\ell + 0.5}{\chi},z \bigg) \, \frac{W^{i}_A(\chi(z)) \, W^{j}_B(\chi(z))}{\chi^2},
\end{align}
where $W^{i}_{A}$ and $W^{j}_{B}$ are the redshift-dependent weight functions for the tomographic bins $i$ and $j$ of probes $A$ and $B$ respectively. For shear of sources in tomographic bin $i$, this corresponds to the lensing efficiency as given by:
\begin{align}\label{eq:Wk}
    W^i_{\kappa}(z) = \frac{3 H_0^2 \Omega_{\rm m}}{2 c^2} \frac{\chi(z)}{a(z)} \int d \chi' n^i_{\kappa}(z') \frac{dz}{d\chi'} \frac{\chi' - \chi}{\chi'} \,
\end{align}
where $n^i_{\kappa}(z')$ is the redshift distribution of the $i^{\rm th}$ tomographic bin of the source galaxy sample (see Fig.~\ref{fig:Wk_nz_tSZ_map}) and $a(z)$ is the scale factor. For tSZ, this weight function is just given by $W_{y}(z) = a(z)$. 

Now, converting these correlations to angular coordinates and using the flat-sky approximation, the tSZ-weak lensing correlation can be calculated using the Hankel transform as
\begin{equation}\label{eq:xigty}
\xi^{i}_{\gamma_t y}(\theta) = \int \frac{d\ell \ \ell}{2\pi} J_{2}(\ell \theta) C^{i}_{\kappa y}(\ell),
\end{equation}
where $J_{2}$ is the second-order Bessel function. Here, $i$ labels the tomographic distribution of source galaxies.

The auto-correlation of shear between any two tomographic bins $i$ and $j$ can be described with two components, $\xi^{ij}_{+}$ and $\xi^{ij}_{-}$, which are given by:
\begin{equation}\label{eq:xiplusminus}
\xi^{ij}_{+/-}(\theta) = \int \frac{d\ell \ \ell}{2\pi} J_{0/4}(\ell \theta) C^{ij}_{\kappa \kappa}(\ell),
\end{equation}
where $J_{0}$ and $J_{4}$ are the zeroth and fourth-order Bessel functions, respectively. Note that while the transform to angular space on a curved sky can be more accurately calculated as detailed in \cite{krause_dark_2021}, we limit our forecast to a maximum angular scale of 250 arcminutes, and thus defer a more precise calculation to future studies.

\subsubsection{Instrinsic alignments and observational systematics}\label{sec:IA}
We assume a simple non-linear alignment model (NLA) to describe the intrinsic alignment (IA) of source galaxies, as suggested by \cite{bridle_dark_2007}. The impact of NLA can be captured by modifying the lensing efficiency as per \cite{krause_dark_2017}:
\begin{equation}
\label{eq:CIA}
W^{i}_{\kappa}(z) \longrightarrow W^{i}_{\kappa}(z) - A\left(z\right) n^i_{\kappa}(z) \frac{dz}{d \chi}\,,    
\end{equation}
where the IA amplitude is modeled using a power-law scaling with amplitude $A_{\rm IA}$ and index $\eta_{\rm IA}$:
\begin{equation}\label{eq:AzIA}
    A(z) = -A_{\rm IA} \bigg(\frac{1+z}{1+z_0}\bigg)^{\eta_{\rm IA}} \frac{C_1 \bar{\rho}_{\rm m,0}}{D(z)},
\end{equation}
and we set $z_0 = 0.62$ and $C_1 = 5\times 10^{-14} M_{\odot}^{-1}h^{-2}{\rm Mpc}^3$ following \cite{brown_measurement_2002}, with $D(z)$ representing the linear growth factor. The parameters $A_{\rm IA}$ and $\eta_{\rm IA}$ are treated as free parameters with the same uninformed wide priors as used in previous DES analyses \citep{secco_dark_2022, amon_dark_2022, des_collaboration_dark_2022} (see Table~\ref{tab:params_all}). Note that the shear$~\times~$tSZ cross-correlation probes higher mass halos where galaxies are more strongly aligned relative to lower mass halos. As shown in \cite{pandey_cross-correlation_2022}, the two lowest-redshift tomographic bins receive contributions on small scales from higher-order intrinsic alignment terms, compared to the NLA model. We exclude the scales below 10 arcmin for the first two tomographic bins of the shear$~\times~$tSZ measurement where the NLA model underestimates the IA signal compared to the predictions from a halo model of IA \citep{fortuna_halo_2021} (see Fig.~3 of \cite{pandey_cross-correlation_2022}). 

We model the photometric uncertainty in our source redshift distribution $n^i_{\kappa}(z)$ using shift parameters ($\Delta^i_{z}$), which modify the source redshift distributions for any tomographic bin $i$ \cite{krause_dark_2017}:
\begin{equation}\label{eq:Delzi}
    n^i_{\kappa}(z) \rightarrow n^i_{\kappa}(z - \Delta^i_{z})
\end{equation}

The multiplicative shear bias modifies the correlations for tomographic bins $i$ and $j$ as follows:
\begin{equation}\label{eq:mi1}
    \xi^{i}_{\gamma_t y}(\theta) \rightarrow (1 + m^i) \, \xi^{i}_{\gamma_t y}(\theta)
\end{equation}
\begin{equation}\label{eq:mi2}
    \xi^{ij}_{+/-}(\theta) \rightarrow (1 + m^i) \, (1 + m^j) \, \xi^{ij}_{+/-}(\theta),
\end{equation}
where $m^i$ represents the multiplicative shear bias parameters. The informative Gaussian priors on  $\Delta^i_{z}$ and $m^i$ are the same as specified in previous DES analyses \citep{secco_dark_2022, amon_dark_2022, des_collaboration_dark_2022} (see Table~\ref{tab:params_all}).

The fiducial DES analysis of the shear two-point auto-correlation \citep{amon_dark_2022, secco_dark_2022}, which used $P_{\rm mm}^{\rm halofit}$ to model the matter power spectrum, removed small-scale measurements to mitigate baryonic feedback effects. These scale cuts were determined for each tomographic bin combination by requiring that the difference between simulated datavectors with and without baryonic effects fell below a designated threshold \citep{secco_dark_2022}. This procedure reduced the measurement's signal-to-noise from 40 (over the full 2.5--250 arcmin range) to 27 \citep{amon_dark_2022}. In contrast, because our model self-consistently incorporates baryonic feedback, we analyze the full data vector across the entire 2.5--250 arcmin range.

\begin{table}
\centering 
\tabcolsep=0.16cm
\begin{tabular}{|c|c c|}
\hline
Parameter &  Prior & Reference  \\ \hline \hline

\multicolumn{3}{|c|}{\textbf{Gas Profile}} \\  \hline
$\theta_{\rm ej, 0}$ & $\mathcal{U}$[1.0, 6.0] & Eq.~\ref{eq:theta_ej_co} \\
$\nu^z_{\theta_{\rm ej}}$ & $\mathcal{U}$[-3.0, 3.0] & Eq.~\ref{eq:theta_ej_co}  \\
$\nu^{M}_{\theta_{\rm ej}}$ & $\mathcal{U}$[-1.0, 1.0] & Eq.~\ref{eq:theta_ej_co} \\  
$\mu_{\beta}$ & $\mathcal{U}$[0.01, 1.5] & Eq.~\ref{eq:beta_gas}  \\
\hline
\hline
\multicolumn{3}{|c|}{\textbf{Non-thermal pressure}} \\ \hline
$\alpha_{\rm nt}$ & $\mathcal{U}$[0.0, 0.5] & Eq.~\ref{eq:Pnt}  \\
\hline
\hline
\multicolumn{3}{|c|}{\textbf{Transition regime}} \\ \hline
$\alpha_{\kappa y}$ & $\mathcal{U}$[0.8, 1.2] & \S~\ref{sec:Pk_xi}  \\
\hline
\hline
\multicolumn{3}{|c|}{\textbf{Cosmology}} \\ \hline
$\Omega_{\rm m}$ & $\mathcal{U}$[0.1, 0.5] &  \\ 
$\Omega_{\rm b}$ & $\mathcal{U}$[0.03, 0.07] &  \\ 
$h$ &$\mathcal{U}$[0.5, 0.9] & \S~\ref{sec:model} \\ 
$n_{\rm s}$ & $\mathcal{U}$[0.8, 1.2] &  \\ 
$\sigma_8$ & $\mathcal{U}$[0.6, 1.0] &  \\ 
\hline
\hline
\multicolumn{3}{|c|}{\textbf{Intrinsic Alignment}} \\ \hline
$A_{\rm IA}$ & $\mathcal{U}$[-5.0, 5.0]  &  Eq.~\ref{eq:AzIA}    \\
$\eta_{\rm IA}$ & $\mathcal{U}$[-5.0, 5.0] &  Eq.~\ref{eq:AzIA}   \\ 
\hline
\hline
\multicolumn{3}{|c|}{\textbf{Shear Calibration}} \\  \hline
$m^{1}$ & $\mathcal{G}[-0.0063, 0.0091]$ & Eq.~\ref{eq:mi1}  \\ 
$m^{2}$ & $\mathcal{G}[-0.0198, 0.0078]$ & Eq.~\ref{eq:mi1}  \\ 
$m^{3}$ & $\mathcal{G}[-0.024, 0.0076]$ & Eq.~\ref{eq:mi1}  \\ 
$m^{4}$ & $\mathcal{G}[-0.037, 0.0076]$ & Eq.~\ref{eq:mi1}  \\ 
\hline
\hline
\multicolumn{3}{|c|}{\textbf{Source photo-$z$ bias}} \\ 
\hline
$\Delta_z^{1}$ & $\mathcal{G}[0.0, 0.018]$ & Eq.~\ref{eq:Delzi}  \\ 
$\Delta_z^{2}$ & $\mathcal{G}[0.0, 0.015]$ & Eq.~\ref{eq:Delzi}  \\ 
$\Delta_z^{3}$ & $\mathcal{G}[0.0, 0.011]$ & Eq.~\ref{eq:Delzi}  \\ 
$\Delta_z^{4}$ & $\mathcal{G}[0.0, 0.017]$ & Eq.~\ref{eq:Delzi}  \\ 
& & \\
\hline 
\end{tabular}
\caption{All the parameters varied in this study along with their priors, specified either as a uniform priors within some minimum and maximum values ($\mathcal{U}({\rm min, max})$ or Gaussian priors with a given mean and standard deviation ($\mathcal{G}({\rm \mu, \sigma})$).}
\label{tab:params_all}
\end{table}

\subsection{Analysis}

All the parameters varied in this analysis, along with their priors, are described in Appendix~\ref{app:all_params}. We assume a standard Gaussian likelihood to sample the parameter space, and the covariance used, along with the sampling methodology, are described in the following.

\subsubsection{Covariance model}\label{sec:covariance}

We model the covariance, $\varmathbb{C}$, as a sum of Gaussian ($\varmathbb{C}^{\rm G}$) and connected non-Gaussian ($\varmathbb{C}^{\rm cNG}$) terms. The multi-probe covariance methodology, including the tSZ observable, is detailed in \cite{fang_cosmology_2024}. We first estimate the covariance in multipole space, employing a methodology similar to that of \cite{fang_cosmology_2024}, but with a few simplifications. We only model the 1-halo part of the connected 4-point function for all probes \citep{friedrich_dark_2021, krause_cosmolike_2017} and ignore the contribution from super-sample covariance \citep{osato_super_2021}. As cross-correlations with the tSZ field are dominated by relatively high-mass halos, we expect the covariance to be dominated by Poisson fluctuations of high-mass objects, as captured by the connected non-Gaussian term \citep{osato_super_2021}. The non-Gaussian covariance contributes at approximately the 10\% level (see Fig.~15 of \cite{gatti_cross-correlation_2022} and \cite{osato_super_2021}). Finally, we convert the covariance from multipole space to angular space, as detailed in \cite{krause_cosmolike_2017, pandey_cross-correlation_2022}. 
The shape noise assumed in the covariance calculation is the same as that presented in \cite{des_collaboration_dark_2022}, and we measure the auto-power spectra of the corresponding tSZ map using the \texttt{NaMaster} package \cite{alonso_unified_2019}. The parameters used for covariance estimation are the best-fit parameters from the first sampling run. We note that the covariance for the shear auto block of the measurement remains identical to that described in \cite{des_collaboration_dark_2022}. 

\subsubsection{Sampling}

As our entire model is developed in the \texttt{JAX} framework, we obtain automatic differentiation of our model’s prediction relative to all the input parameters. This means that we can calculate the gradient of the posterior at any point relative to the parameters, which can be used in gradient-based sampling. This is advantageous because it makes the sampling significantly more efficient. We use the Hamiltonian Monte Carlo (HMC) method to sample the parameter space \citep{duane_hybrid_1987}. This sampling method is modeled analogously to a physical system, where the total Hamiltonian of the system is defined as the potential energy (the negative logarithm of the posterior) plus the kinetic energy, which is obtained by defining momentum variables for each parameter in the sampling space. This allows the sampling to traverse along the directions of constant energy, resulting in a high probability of acceptance for the next proposed state. The sampling efficiency remains constant (approximately 70\%), even in large dimensions, which is significantly higher than that of traditional MCMC algorithms (see \cite{neal_mcmc_2011, betancourt_conceptual_2018} for a review of HMC methods). Note that the HMC sampling requires specifying the step size and the number of leapfrog steps, which vary significantly based on the number of dimensions and the shape of the posterior. This limits the out-of-the-box applicability of the standard HMC algorithm to general problems, as it requires tuning these two parameters.

The No-U-Turn Sampler (\texttt{NUTS}), first described in \cite{hoffman_no-u-turn_2011}, circumvents this issue by adaptively determining these parameters for the problem at hand. This method dynamically tunes the step size of the sampler based on its trajectory in the sampling space. This method adjusts the step size such that the average gradient of the log posterior scales appropriately relative to its curvature. Moreover, this method builds a binary tree of states (up to a maximum of $2^{\rm t_{\rm depth}}$ steps) as the sampler explores the space and evaluates whether it is feasible to continue in the current direction. This decision is made using the ``no-u-turn” criterion, which checks whether the sampler is returning to the starting point by doubling back. This method adaptively sets the number of leapfrog steps and significantly improves the efficiency of the sampler for any general problem. These sampling methods have been implemented in cosmological settings such as in CMB data analysis \citep{hajian_efficient_2007, taylor_fast_2008}, initial condition reconstruction \citep{jasche_physical_2019}, and LSS correlation analysis \citep{campagne_jax-cosmo_2023, mootoovaloo_assessment_2024, piras_future_2024}. 

We use the implementation of \texttt{NUTS} available in the \texttt{numpyro} library.\footnote{\url{https://num.pyro.ai/en/latest/index.html}} We run 64 chains in parallel that are initialized at random positions close to the median of the prior on 4 \texttt{Nvidia-H100} GPUs. We set an initial step size of 0.3 and $t_{\rm depth} = 4$ for computational efficiency and verify that our constraints do not change when increasing $t_{\rm depth}$ or changing the initial step size. We then run 8000 samples in the warm-up phase for each chain, during which we adapt the mass matrix and step sizes, which are then fixed during the sampling phase. Finally, we obtain 8000 samples from each of the 64 chains, which we use to analyze our results. This entire process takes approximately 6 hours. We ensure that our samples satisfy the convergence criteria of $R-1 < 0.01 $ where $R$ is the Gelman-Rubin statistic \citep{gelman_inference_1992}. In Appendix~\ref{app:code_val}, we show a comparison of the chain run with our default \texttt{NUTS} implementation, compared to the standard nested sampling approach \citep{handley_polychord_2015}, as used in the fiducial DES Y3 analysis \citep{des_collaboration_dark_2022}. Several improvements are possible for the \texttt{NUTS} sampling to make it even more efficient, as detailed in \cite{campagne_jax-cosmo_2023}, which we leave for future exploration.

\begin{figure}
\centering
\includegraphics[width=0.49\textwidth]{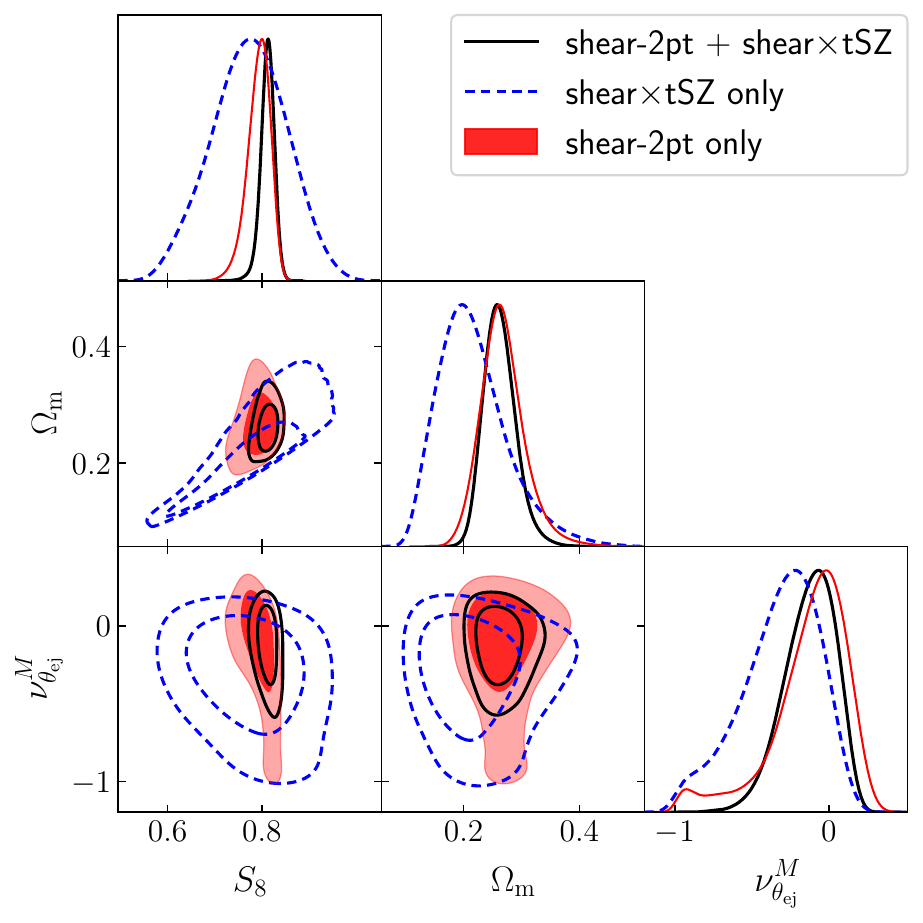}
\caption[]{Constraints on two of the cosmological parameters ($\Omega_{\rm m}$ and $S_8 = \sigma_8 (\Omega_{\rm m}/0.3)^{0.5}$) and one baryonic parameter $\nu^M_{\theta_{\rm ej}}$ (see Eq.~\ref{eq:theta_ej_co}) controlling the evolution of the gas ejection radius with halo mass, when individually analyzing either the shear auto-correlation (solid red) or shear$~\times~$tSZ cross-correlation (dashed blue) and when jointly analyzing both probes with our model (thin black).}
\label{fig:Om_S8_muej_internal}
\end{figure}

\begin{figure}
\centering
\includegraphics[width=0.49\textwidth]{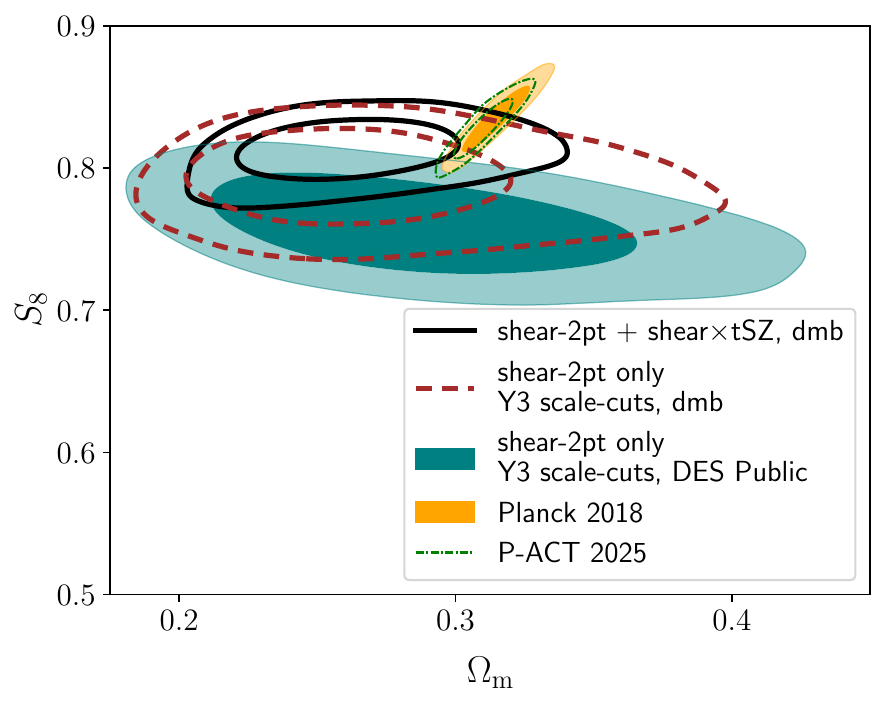}
\caption[]{We compare the constraints obtained from our joint analysis of shear auto or shear$~\times~$tSZ (black, unfilled contours) with the public constraints from DES (teal, filled contours) \cite{amon_dark_2022, secco_dark_2022} analyzing the large-scale shear auto-correlation and from the \planck\ as well as \planck\ + ACT joint CMB analysis (orange, filled contours and green unfilled contours respectively) \cite{collaboration_planck_2020, louis_atacama_2025}. In the brown dashed unfilled contours, we additionally show the constraints obtained from our model when analyzing only the large-scale shear auto-correlation.}
\label{fig:Om_S8_external}
\end{figure}

\begin{figure}
\centering
\includegraphics[width=0.49\textwidth]{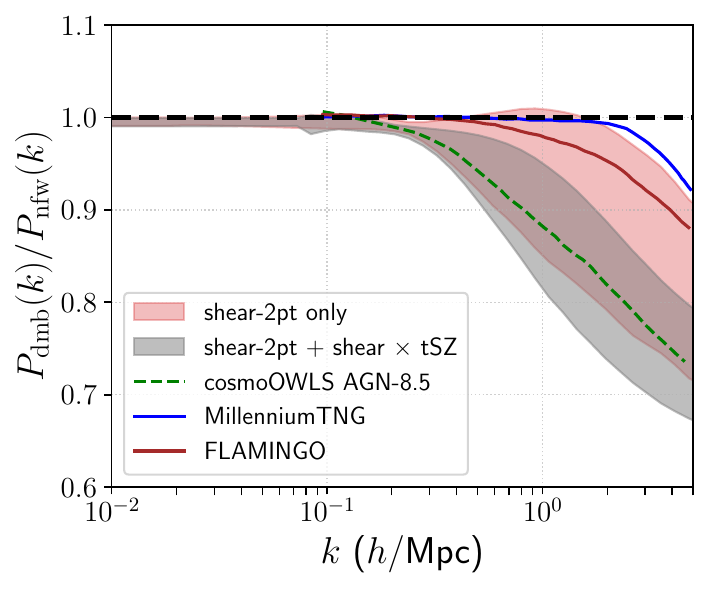}
\caption[]{Inferred 1$\sigma$ constraints on the matter power suppression at $z=0$. We show the constraints obtained from analysis of just the shear auto-correlation, as well as from the joint analysis of shear auto and shear$~\times~$tSZ. In solid and dashed lines, the power suppression predicted from various simulations is also shown.}
\label{fig:DP_P}
\end{figure}

\section{Results}
\label{sec:results}

We show the subset of cosmological and baryonic parameter constraints obtained from analyzing the shear auto-correlation and shear$~\times~$tSZ cross-correlation as individual probes, as well as from their joint analysis, in Fig.~\ref{fig:Om_S8_muej_internal}. We focus on two cosmological parameters, the total matter density $\Omega_{\rm m}$ and the late-time matter clustering amplitude $S_8 = \sigma_8 (\Omega_{\rm m}/0.3)^{0.5}$ that the shear auto-correlation is sensitive to. Note that we vary five cosmological parameters in our analysis with a broad uniform prior, as shown in Appendix~\ref{app:all_params}. We see that as expected, shear auto provides better constraints on the cosmological parameters compared to shear$~\times~$tSZ. However, their joint analysis breaks the degeneracies in the parameter space, leading to tighter constraints. The figure of merit in the $\Omega_{\rm m}-S_8$ plane of the constraints improves from 1062 when analyzing the shear two-point auto-correlation individually to 2454 in the case of joint analysis of the shear two-point auto and shear$~\times~$tSZ cross-correlation. From the analysis of the shear two-point only, we obtain $S_8 = 0.793^{+0.028}_{-0.019}$ and $\Omega_{\rm m} = 0.268^{+0.030}_{-0.041}$ , whereas with the joint analysis, we obtain $S_8 = 0.811^{+0.015}_{-0.012}$ and $\Omega_{\rm m} = 0.263^{+0.023}_{-0.030}$. We also show one baryonic parameter, $\nu_{\theta_{\rm ej}}^{M}$ (see Eq.~\ref{eq:theta_ej_co}). which controls the evolution of the gas ejection radius as a function of halo mass. A negative value indicates that gas is ejected out further in lower-mass halos than in higher-mass halos, as expected due to their shallower potential wells. We find that the data weakly prefer negative  $\nu_{\theta_{\rm ej}}^{M}$, with more precise constraints obtained in the case of the joint analysis ($\nu_{\theta_{\rm ej}}^{M} = -0.124^{+0.194}_{-0.132}$) \citep{pandey_cross-correlation_2022}. A plot with constraints on all the parameters varied in the analysis is shown in Appendix~\ref{app:all_params}. 

In Fig.~\ref{fig:Om_S8_external}, we compare the cosmological constraints obtained here from joint shear and tSZ analysis with those from other studies. In filled teal contours we show the fiducial DES-Y3 public constraints as obtained from large-scale analysis of shear auto-correlations \cite{amon_dark_2022, secco_dark_2022}, which uses a dark-matter only model for cosmic shear. In addition, we also show the constraints obtained from the primary CMB analysis of the observations from the \planck\, satellite \citep{collaboration_planck_2020} as well as joint CMB analysis from the \planck\ and ACT DR6 CMB observations as presented in \cite{louis_atacama_2025}. We see that our constraints are tighter compared to the DES public results and consistent with CMB constraints. These constraints are also consistent with cosmological results obtained by analyzing galaxy clustering data with Dark Energy Spectroscopic Instrument (DESI, \citep{collaboration_desi_2024}) within the $\Lambda$CDM model. We additionally show the constraints obtained by analyzing the large-scale shear auto data with the same scale cuts as used in the DES public results, but with our full model for the matter distribution and analysis pipeline in the brown dashed contours (marginalizing over the parameters listed in Table~\ref{tab:params_all}). The upward shift in the $S_8$ value, in both the shear auto only and joint contours from this work, compared to the DES public results, is caused by switching to an alternative model of matter clustering that accounts for the effects of baryonic feedback (similar effects are described in \cite{secco_dark_2022}) and are also seen in other recent studies reanalyzing DES data \citep{bigwood_weak_2024, anbajagane_decade_2025, arico_y3_2023, DarkEnergySurveyandKilo-DegreeSurveyCollaboration:2023:OJAp:}. The different model of intrinsic alignments and excluding the likelihood of shear ratios in our pipeline also contribute to this shift, as described in \cite{secco_dark_2022, arico_y3_2023, doux_dark_2022}. We leave the extension of our model to include more flexible intrinsic alignment models \citep{blazek_tidal_2015, fortuna_halo_2021} as well as more cosmological models, such as $w_0-w_a$CDM and massive neutrinos to future study.

We show the best-fit theory predictions from the joint analysis overlaid on the measurements of shear$~\times~$tSZ in Fig.~\ref{fig:xigty_measure} and for shear auto in Fig.~\ref{fig:xip_measure} and Fig.~\ref{fig:xim_measure}. The best-fit $\chi^2$ is 510.2 for 468 data points. Out of all the parameters varied (21 parameters), approximately 8 were constrained relative to their prior as estimated from the \texttt{tensiometer} package.\footnote{https://github.com/mraveri/tensiometer} This results in a $\chi^2$ per degree of freedom of approximately 1.1 and a $p-$value of approximately 0.072, indicating a good fit.

We use the parameter constraints obtained on our model to predict the matter power suppression due to baryonic effects relative to a dark matter only prediction, $P_{\rm dmb}/P_{\rm nfw}$ at $z=0$ as a function of scale. By evaluating this quantity for 1000 random points from the parameter posterior, we extract the $16^{\rm th}$ and $84^{\rm th}$ percentile constraint as shown in Fig.~\ref{fig:DP_P}. We show the matter power suppression as obtained from just the analysis of shear auto data as well as from the joint analysis with shear$~\times~$tSZ. Additionally, we also show the prediction of the power suppression from various hydrodynamical simulations \citep{van_daalen_impact_2014, hernandez-aguayo_millenniumtng_2023, schaye_flamingo_2023}. Due to different subgrid physics implementations as well as assumptions about the strength and mechanisms of baryonic feedback, these simulations predict characteristically different scale dependence of the power suppression. We find that our joint analysis predicts larger power suppression relative to the Millennium TNG \citep{hernandez-aguayo_millenniumtng_2023} simulation and the fiducial \texttt{Flamingo} simulation \citep{schaye_flamingo_2023}, and agrees remarkably well with the AGN-8.5 version of the cosmo-OWLS simulations as described in \cite{van_daalen_impact_2014, le_brun_towards_2014}, which increases the temperature of a subset of gas particles to $\Delta T_{\rm heat} = 3 \times 10^8 \, {\rm K}$, mimicking the effect of increased AGN feedback. 

To quantify the tension between the simulation curves and the inferred matter power suppression constraints, we calculate the reduced $\Delta \chi^2$ between the inferred suppression at six logarithmically spaced $k-$values between $0.01 \, h/{\rm Mpc}$ and $5 \, h/{\rm Mpc}$ and predictions from the three hydro simulations shown in Fig.~\ref{fig:DP_P}. We use the power suppression obtained at 10000 random samples to get the covariance which accounts for correlations between the scales and we convert the reduced $\Delta \chi^2$ to significance of tension.
We find that the OWLS-AGN 8.5 \citep{van_daalen_impact_2014}, fiducial \texttt{FLAMINGO} \citep{schaye_flamingo_2023} and \texttt{MTNG} \citep{hernandez-aguayo_millenniumtng_2023} simulations are at 0.3$\sigma$, 2.3$\sigma$ and $3.8 \sigma$ deviation relative to our inferred suppression.
These constraints on matter power suppression are consistent with previous observational analyses that include the small-scale shear auto-correlation \citep{chen_constraining_2023}, include the prior from kSZ studies \citep{amodeo_atacama_2021, bigwood_weak_2024}, jointly analyze shear and kSZ with X-ray data \citep{schneider_constraining_2022}, or use a large library of simulations as an emulator for the tSZ effect \citep{pandey_inferring_2023}.

\begin{figure}
\centering
\includegraphics[width=0.49\textwidth]{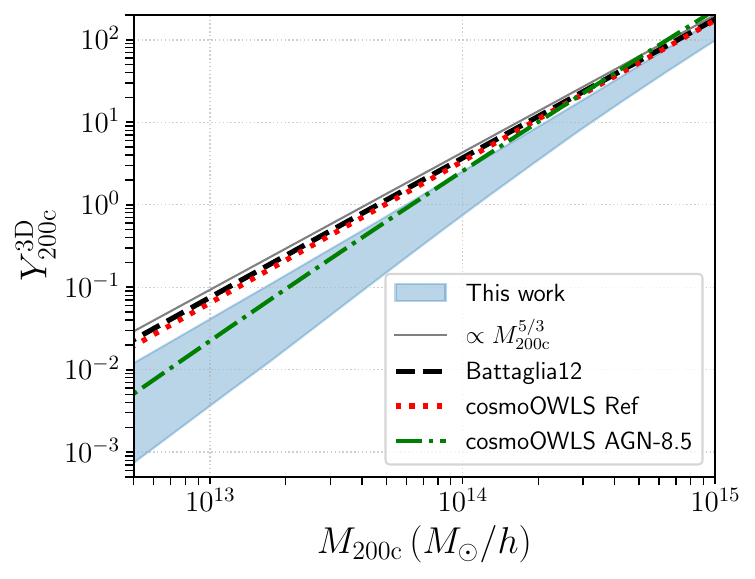}
\caption[]{Inferred $1\sigma$ constraints on the $Y^{\rm 3D}_{\rm 200c}-M_{\rm 200c}$ relationship obtained from the joint shear auto + shear$~\times~$tSZ analysis (shaded blue) and its comparison with different hydrodynamical simulations (dashed and dotted lines), as well as the self-similar $Y^{\rm 3D}_{\rm 200c} \propto M_{\rm 200c}^{5/3}$ power-law scaling relation (thin solid line).}
\label{fig:YM}
\end{figure}

\begin{figure}
\centering
\includegraphics[width=0.49\textwidth]{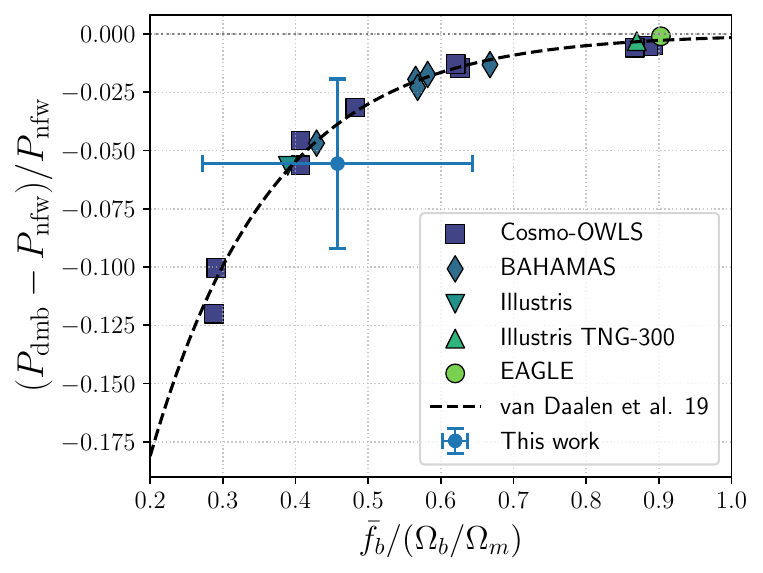}
\caption[]{Inferred constraints on the matter power suppression at $k = 0.5 \, h/{\rm Mpc}$ and the baryon fraction in halos with mass $M_{\rm 500c}~\sim~10^{14} M_{\odot}$. Our constraints are compared with results from various hydrodynamical simulations that implement different baryonic feedback prescriptions, as well as the best-fit curve derived from the measurements, as detailed in \cite{daalen_exploring_2020}.
}
\label{fig:vd19_comp}
\end{figure}

In Fig.~\ref{fig:YM}, we use the parameter constraints from the jointly fit model to infer the $16^{\rm th}$ and $84^{\rm th}$ percentile constraints on the pressure integrated within the halo radius: $Y^{\rm 3D}_{\rm 200c}(M) = \int_0^{r_{\rm 200c}} dr \, 4\pi r^2 \, P_e(M,r)$ at $z=0$.  We additionally over-plot the predictions from various hydrodynamical simulations \citep{van_daalen_effects_2011, battaglia_cluster_2012}. In an isolated gravitationally bound system without any baryonic feedback one would predict $Y^{\rm 3D}_{\rm 200c} \propto M_{\rm 200c}^{5/3}$ \citep{kaiser_evolution_1991, nagai_effects_2007}. The universal pressure profiles obtained by fitting to high-mass X-ray clusters have also shown very similar scaling \citep{arnaud_universal_2010}. Similarly, the simulations of \citep{battaglia_simulations_2010} as well as the reference model of \cite{le_brun_towards_2014}, which have a relatively mild AGN feedback prescription, follow this scaling down to low masses. However, recently various studies have shown a break in the $Y-M$ power-law relation, particularly for group-scale halo masses ($M \lesssim 10^{14} \, M_{\odot}/h$), with lower-mass halos preferring a steeper slope \citep{greco_stacked_2015, hill_two-halo_2018, pandey_cross-correlation_2022, osato_investigating_2018}. This occurs due to stronger baryonic feedback, particularly from AGN, which blows away the gas from inside the halos, causing a reduction in the effective baryonic thermal energy inside the halo radius. This effect gets stronger for lower-mass halos, which have shallower potential wells, making it easier for the gas to be blown out. We see that our inferred constraints also show a similar trend, preferring a steeper slope in the $Y-M$ relation and again being consistent with the AGN-8.5 suite of cosmo-OWLS simulations \citep{le_brun_towards_2014, le_brun_testing_2015}. We again quantify the tension between the hydrodynamical simulation predictions and our inference by calculating the reduced $\Delta \chi^2$ between our $Y-M $ relation inference and the simulation predictions shown in Fig.~\ref{fig:YM}. Note that here we  neglect any stochastic error contribution to the calculated $Y-M $ relation from the simulations.
We find that the simulations of \citep{battaglia_cluster_2012} and the OWLS-ref simulation of \citep{le_brun_towards_2014} deviate mildly, at 1.9$\sigma$ and $1.8 \sigma$ respectively, relative to our inferred constraints. This is consistent with similar findings from previous studies as well \citep{pop_sunyaev-zeldovich_2022, hadzhiyska_interpreting_2023, wadekar_augmenting_2023, battaglia_deconstructing_2015}. Additionally, we also verify that the inferred tSZ auto-power spectra from our parameter constraints is consistent with results presented in \cite{louis_atacama_2025, reichardt_improved_2021, bolliet_dark_2018}. 

As shown in both Fig.~\ref{fig:DP_P} and Fig.~\ref{fig:YM}, we see correlated impacts of increased baryonic feedback on both the matter density and baryonic thermodynamics. A similar correlated observable was identified in \cite{daalen_exploring_2020} using a large suite of hydrodynamical simulations varying a wide range of feedback phenomena, showing that the baryon fraction of halos with mass $M \sim 10^{14} M_{\odot}$ is highly correlated with the matter power suppression at $k \sim 0.5 \, h/{\rm Mpc}$ (also see \cite{delgado_predicting_2023, pandey_inferring_2023, salcido_spk_2023}). We use the inferred parameter constraints from our joint analysis to obtain the mean baryon fraction within $R_{\rm 500c}$ of halos in the same range as used in \cite{daalen_exploring_2020}, $6\times 10^{13} < M_{\rm 500c} [M_{\odot}] < 2 \times 10^{14}$. In particular, using 1000 randomly selected parameters sets from our converged analysis chain, we calculate the gas mass and total mass inside $r_{\rm 200c}$ of halos, $M_{\rm gas/dmb}(M_{\rm 200c}) = \int_0^{r_{\rm 200c}} 4\pi \, r^2 \, \rho_{\rm gas/dmb}(r|M_{\rm 200c})$ for each set, which is then used to estimate the baryon fraction $f_b(M_{\rm 200c}) = [M_{\rm gas}(M_{\rm 200c})]/[M_{\rm dmb}(M_{\rm 200c})]$. These samples are then integrated over the halo mass range of interest to obtain the mean baryon fraction samples: 
\begin{equation}
    \bar{f_b} = \frac{\int_{6e13}^{2e14} dM_{\rm 200c} \, \frac{dn}{dM} \, f_b(M_{\rm 200c})}{\int_{6e13}^{2e14} dM_{\rm 200c} \, \frac{dn}{dM}}.
\end{equation}
We then plot the mean and 1$\sigma$ error on both the mean baryon fraction relative to the cosmic baryon fraction ($\bar{f_b}/(\Omega_b/\Omega_m)$) and the matter power suppression at $k = 0.5 \, h/{\rm Mpc}$ (see Fig.~\ref{fig:DP_P}) in Fig.~\ref{fig:vd19_comp}. We over-plot the prediction of these variables from a large suite of hydrodynamical simulations, as well as the fitting function described in \cite{daalen_exploring_2020}, finding that our inferred constraints are remarkably consistent with the trend seen in the simulations. The constraints on the baryon fraction are also roughly consistent with X-ray observations (e.g. \cite{Budzynski2014, Kravtsov2018, Pearson2017}). We also agree with the high AGN variant of the cosmo-OWLS simulations \cite{le_brun_towards_2014} as well as the original Illustris simulation \cite{vogelsberger_introducing_2014}. Recently, a similar conclusion was reached by analyzing the kSZ signal around the galaxies as described in \cite{hadzhiyska_evidence_2024, mccarthy_flamingo_2025}. Note that some of these simulations have trouble reproducing other observables such as galaxy colors and morphologies \citep{kauffmann_morphology_2019, rodriguez-gomez_optical_2019}, perhaps reflecting a need to further refine the physical implementation of the feedback processes.

\section{Discussion}
\label{sec:conclusion}

In this paper, we detected and analyzed the cross-correlation between the tSZ effect measured from ACT+\textit{Planck} and weak lensing of galaxies as measured from the first three years of observations from DES.  We detected the cross-correlation signal at 21$\sigma$.  We analyzed this correlation jointly with the auto-correlation for weak lensing (shear auto) as measured by DES Y3  \citep{amon_dark_2022, secco_dark_2022}. This is enabled by an accurate joint model of the baryonic thermodynamics and the matter distribution that is validated on a large suite of hydrodynamical simulations \citep{pandey_accurate_2025}. Moreover, this analysis is sensitive to the interesting halo mass range of $5 \lesssim 10^{13} \lesssim M (M_{\odot}/h) \lesssim 5 \times 10^{14}$ and halo redshift range of $0.2 \lesssim z \lesssim 0.7$ and probes the matter power spectrum on scales $k \lesssim 5 \, h/{\rm Mpc}$. 

We find that, as expected, a joint analysis of both the shear auto and shear$~\times~$tSZ probes leads to degeneracy-breaking in our parameter space, leading to tighter cosmological and astrophysical constraints. We find that our inferred cosmological constraints are consistent with the \planck\, and P-ACT primary CMB constraints. Moreover, we find clear signatures of increased baryonic feedback, particularly for the lower halo masses ($M < 10^{14} \, M_{\odot}/h$). This could be caused by increased AGN effects due to shallower potential wells of these group mass halos, ejecting the gas out of the halo boundary. Our inferred constraints on the suppression of matter power, integrated tSZ signal, and integrated baryon fraction within the halo radius are consistent with hydrodynamical simulations that have strong AGN feedback. When comparing with various hydro simulations, we find that some simulations with mild AGN feedback are at more than $4\sigma$ tension with our inferences. However, note that simulations with strong AGN feedback struggle in reproducing other observables such as galaxy colors and their morphologies as well as X-ray observations in clusters \citep{le_brun_towards_2014}.

This opens up two interesting routes to develop better physical understanding. The first is to develop a joint model of galaxy properties and gas thermodynamics. This can be achieved either with a large suite of hydrodynamical simulations with varying feedback prescriptions and their strengths \citep{villaescusa-navarro_camels_2021, Bigwood:2025:arXiv:} at sufficient volume, or by developing effective physical models \citep{pandya_unified_2023}. On the other hand, there could be contributions from other poorly understood sources impacting the ionized free electrons that are not accurately captured in the current simulations' framework. Recently, \cite{hopkins_cosmic_2025, quataert_cosmic_2025, su_modeling_2025} have shown that cosmic rays generated from high energy events (such as AGN and supernovae) can couple to the ionized electrons, causing similar features in the matter power suppression and integrated tSZ and baryon fraction that we see here. It would be important to include their contributions in the upcoming cosmological hydrodynamical simulations to better understand their impact as a function of halo mass, redshift, and environment. 

Another route to robustly understand baryonic feedback is to include more probes of gas thermodynamics, such as X-ray and kSZ \cite{battaglia_future_2017, amodeo_atacama_2021,la_posta_xy_2024}. As these probes are sensitive to different halo masses and redshifts, performing a joint analysis with lensing and tSZ would be able to test the modeling on a large dynamical range of halo masses and redshifts. As next-generation observatories will start generating data in this decade \citep{collaboration_lsst_2021, collaboration_simons_2019}, it is imperative to develop better models of the components of the large-scale structure to develop a deeper physical understanding of multi-wavelength, multi-probe observations of our Universe.

\section*{Acknowledgements}
This paper has gone through internal review by the DES and ACT collaborations. 
CS acknowledges support from the Agencia Nacional de Investigaci\'on y Desarrollo (ANID) through Basal project FB210003.  JCH acknowledges support from NSF grant AST-2108536, the Sloan Foundation, and the Simons Foundation.

Funding for the DES Projects has been provided by the U.S. Department of Energy, the U.S. National Science Foundation, the Ministry of Science and Education of Spain, the Science and Technology Facilities Council of the United Kingdom, the Higher Education Funding Council for England, the National Center for Supercomputing Applications at the University of Illinois at Urbana-Champaign, the Kavli Institute of Cosmological Physics at the University of Chicago, the Center for Cosmology and Astro-Particle Physics at the Ohio State University, the Mitchell Institute for Fundamental Physics and Astronomy at Texas A\&M University, Financiadora de Estudos e Projetos, Fundação Carlos Chagas Filho de Amparo a Pesquisa do Estado do Rio de Janeiro, Conselho Nacional de Desenvolvimento Científico e Tecnológico and the Ministério da Ciência, Tecnologia e Inovação, the Deutsche Forschungsgemeinschaft and the Collaborating Institutions in the Dark Energy Survey.

The Collaborating Institutions are Argonne National Laboratory, the University of California at Santa Cruz, the University of Cambridge, Centro de Investigaciones Energeticas, Medioambientales y Tecnológicas-Madrid, the University of Chicago, University College London, the DES-Brazil Consortium, the University of Edinburgh, the Eidgenossische Technische Hochschule (ETH) Zurich, Fermi National Accelerator Laboratory, the University of Illinois at Urbana-Champaign, the Institut de Ciencies de l'Espai (IEEC/CSIC), the Institut de Fisica d'Altes Energies, Lawrence Berkeley National Laboratory, the Ludwig-Maximilians Universität München and the associated Excellence Cluster Universe, the University of Michigan, NSF NOIRLab, the University of Nottingham, The Ohio State University, the University of Pennsylvania, the University of Portsmouth, SLAC National Accelerator Laboratory, Stanford University, the University of Sussex, Texas A\&M University, and the OzDES Membership Consortium.

Based in part on observations at NSF Cerro Tololo Inter-American Observatory, NSF NOIRLab (NOIRLab Prop. ID 2012B-0001; PI: J. Frieman), which is managed by the Association of Universities for Research in Astronomy (AURA) under a cooperative agreement with the U.S. National Science Foundation.

The DES data management system is supported by the National Science Foundation under Grant Numbers AST-1138766 and AST-1536171.

The DES participants from Spanish institutions are partially supported by MICINN under grants PID2021-123012, PID2021-128989 PID2022-141079, SEV-2016-0588, CEX2020-001058-M and CEX2020-001007-S, some of which include ERDF funds from the European Union. IFAE is partially funded by the CERCA program of the Generalitat de Catalunya.
We acknowledge support from the Brazilian Instituto Nacional de Ciencia
e Tecnologia (INCT) do e-Universo (CNPq grant 465376/2014-2).

This document was prepared by the DES Collaboration using the resources of the Fermi National Accelerator Laboratory (Fermilab), a U.S. Department of Energy, Office of Science, Office of High Energy Physics HEP User Facility. Fermilab is managed by Fermi Forward Discovery Group, LLC, acting under Contract No. 89243024CSC000002.

Here is the same text in LaTex format (cut-and-paste), see version history for last update.

Funding for the DES Projects has been provided by the U.S. Department of Energy, the U.S. National Science Foundation, the Ministry of Science and Education of Spain, 
the Science and Technology Facilities Council of the United Kingdom, the Higher Education Funding Council for England, the National Center for Supercomputing 
Applications at the University of Illinois at Urbana-Champaign, the Kavli Institute of Cosmological Physics at the University of Chicago, 
the Center for Cosmology and Astro-Particle Physics at the Ohio State University,
the Mitchell Institute for Fundamental Physics and Astronomy at Texas A\&M University, Financiadora de Estudos e Projetos, 
Funda{\c c}{\~a}o Carlos Chagas Filho de Amparo {\`a} Pesquisa do Estado do Rio de Janeiro, Conselho Nacional de Desenvolvimento Cient{\'i}fico e Tecnol{\'o}gico and 
the Minist{\'e}rio da Ci{\^e}ncia, Tecnologia e Inova{\c c}{\~a}o, the Deutsche Forschungsgemeinschaft and the Collaborating Institutions in the Dark Energy Survey. 

The Collaborating Institutions are Argonne National Laboratory, the University of California at Santa Cruz, the University of Cambridge, Centro de Investigaciones Energ{\'e}ticas, 
Medioambientales y Tecnol{\'o}gicas-Madrid, the University of Chicago, University College London, the DES-Brazil Consortium, the University of Edinburgh, 
the Eidgen{\"o}ssische Technische Hochschule (ETH) Z{\"u}rich, 
Fermi National Accelerator Laboratory, the University of Illinois at Urbana-Champaign, the Institut de Ci{\`e}ncies de l'Espai (IEEC/CSIC), 
the Institut de F{\'i}sica d'Altes Energies, Lawrence Berkeley National Laboratory, the Ludwig-Maximilians Universit{\"a}t M{\"u}nchen and the associated Excellence Cluster Universe, 
the University of Michigan, NSF NOIRLab, the University of Nottingham, The Ohio State University, the University of Pennsylvania, the University of Portsmouth, 
SLAC National Accelerator Laboratory, Stanford University, the University of Sussex, Texas A\&M University, and the OzDES Membership Consortium.

Based in part on observations at NSF Cerro Tololo Inter-American Observatory at NSF NOIRLab (NOIRLab Prop. ID 2012B-0001; PI: J. Frieman), which is managed by the Association of Universities for Research in Astronomy (AURA) under a cooperative agreement with the National Science Foundation.

The DES data management system is supported by the National Science Foundation under Grant Numbers AST-1138766 and AST-1536171.
The DES participants from Spanish institutions are partially supported by MICINN under grants PID2021-123012, PID2021-128989 PID2022-141079, SEV-2016-0588, CEX2020-001058-M and CEX2020-001007-S, some of which include ERDF funds from the European Union. IFAE is partially funded by the CERCA program of the Generalitat de Catalunya.

We  acknowledge support from the Brazilian Instituto Nacional de Ci\^encia
e Tecnologia (INCT) do e-Universo (CNPq grant 465376/2014-2).

This document was prepared by the DES Collaboration using the resources of the Fermi National Accelerator Laboratory (Fermilab), a U.S. Department of Energy, Office of Science, Office of High Energy Physics HEP User Facility. Fermilab is managed by Fermi Forward Discovery Group, LLC, acting under Contract No. 89243024CSC000002.

Support for ACT was through the U.S. National Science Foundation through awards AST-0408698, AST0965625, and AST-1440226 for the ACT project, as well as awards PHY-0355328, PHY-0855887 and PHY1214379. Funding was also provided by Princeton University, the University of Pennsylvania, and a Canada Foundation for Innovation (CFI) award to UBC. ACT
operated in the Parque Astron´omico Atacama in northern Chile under the auspices of the Agencia Nacional de Investigaci´on y Desarrollo (ANID). The development of multichroic detectors and lenses was supported by NASA grants NNX13AE56G and NNX14AB58G. Detector research at NIST was supported by the NIST
Innovations in Measurement Science program. This research also used resources of the National Energy Research Scientific Computing Center (NERSC), a U.S. Department of Energy Office of Science User Facility located at Lawrence Berkeley National Laboratory, operated under Contract No. DE-AC02-05CH11231 using NERSC award HEPERCAPmp107 from 2021 to 2025. We thank the Republic of Chile for hosting ACT in the northern Atacama, and the local indigenous Licanantay communities whom we follow in observing and learning from the night sky.

\bibliographystyle{mnras}
\bibliography{ads}

\newpage
\appendix

\section{Impact of systematics}
\label{app:cib}

\subsection{Point sources}
In the top row of Fig.~\ref{fig:PS_cib_deproj_variations}, we show the shear$~\times~$tSZ measurements when using different masks for point sources (primarily bright AGN). With circle markers we show the measurements as obtained when using the public ACT + \emph{Planck} tSZ y-map with $\beta + d\beta$ deprojected (see below). This map is constructed by subtracting the contribution from the point sources and inpainting the extended sources in the frequency maps. The point sources detected at more than $5\sigma$ were subtracted, sources that are detected at more than $70\sigma$ were inpainted with a hole of 6 arcmin, and larger non-SZ extended sources were inpainted with a hole of 10 arcmin radius \citep{naess_atacama_2025, coulton_atacama_2024}.
With square markers, we show the results when we completely mask out both these extended sources ($\sim 1000$ objects) that are inpainted. We see that this does not change our measurements. With diamond markers we show our fiducial measurement when we in addition also mask out all the point sources ($\sim 20000$ objects) with a 3 arcmin hole mask finding that this slightly changes our measurements, particularly at higher redshift bins. Additionally, with inverted-V markers we show the results when changing the hole size to 4.5 arcmin around each point source, finding consistent results. 

\begin{figure}
\centering
\includegraphics[width=0.49\textwidth]{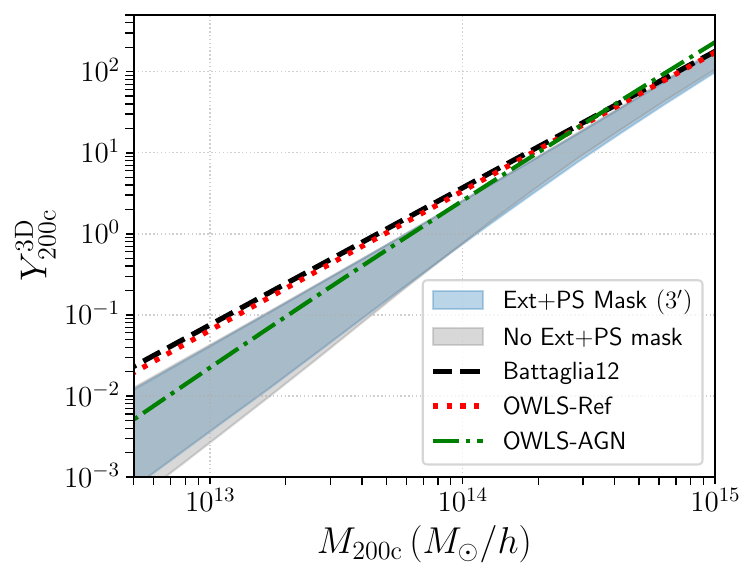}
\caption[]{Robustness of the $Y-M$ relationship to either masking out the point and extended sources detected in ACT (blue region) or ignoring such a mask (gray region). See Appendix~\ref{app:cib} for details.}
\label{fig:YM_PS_mask}
\end{figure}

We will show in Fig.~\ref{fig:allparams} that our constraints on all the cosmological parameters remain insensitive to the choice of point source mask. Moreover, in Fig.~\ref{fig:YM_PS_mask} we show the comparison of our inferred $Y-M$ relation when including or excluding the mask around the point and extended sources and find that our astrophysical inferences are also robust to this choice. We leave investigating the impact of unresolved radio sources on our measurements to a future study.

\subsection{CIB}

The microwave sky observations ($T_{\nu}(\vec{n})$)  at any frequency $\nu$ and direction $\vec{n}$ can be decomposed into individual components, such as tSZ ($y_{\nu}(\vec{n})$), CIB ($C_{\nu}(\vec{n})$), noise ($N_{\nu}(\vec{n})$), and contributions from other residual components ($R_{\nu}(\vec{n})$). We theoretically know the tSZ SED ($f_{\nu}$) \citep{sunyaev_small-scale_1970}, hence we can write $y_{\nu}(\vec{n}) = f_{\nu} y(\vec{n})$, where
\begin{equation}
    f_{\nu} = x \, {\rm coth}(x/2) - 4,
\end{equation}
where ${x = h\nu/k_{\rm B} T_{\rm CMB}}$, $h$ is Planck's constant, $k_{\rm B}$ is Boltzmann's constant, and $T_{\rm CMB}=2.726\, {\rm K}$ is the mean CMB temperature. We can also assume that the CIB can be modeled as an effective modified blackbody with SED $g_{\nu}$:
\begin{equation}
    g_{\nu} = \frac{A \left(\frac{\nu}{\nu_0}\right)^{3+\beta}}{\exp \left(\frac{h\nu}{k_B T_{\rm CIB}}\right)-1} \left(\frac{dB(\nu, T)}{dT}\biggr\rvert_{T = T_{\mathrm{CMB}}}\right)^{-1} \;,
\end{equation}
where $B(\nu, T)$ is the Planck function, $\beta = 1.7$, $T_{\rm CIB} = 10.70$, $A$ is the normalization constant, and $\nu_0$ is the normalization frequency. These parameters are obtained by fitting CIB monopole measurements as described in \citep{mccarthy_component-separated_2024}. The multi-frequency observations from ACT and \textit{Planck} can then be combined to obtain a map that gives unit response to $y$-component with SED of $f_{\nu}$ and zero response to CIB component with SED of $g_{\nu}$.

\begin{figure*}
    \centering
    \vfill
    \subfloat[]{
        \includegraphics[width=0.99\linewidth]{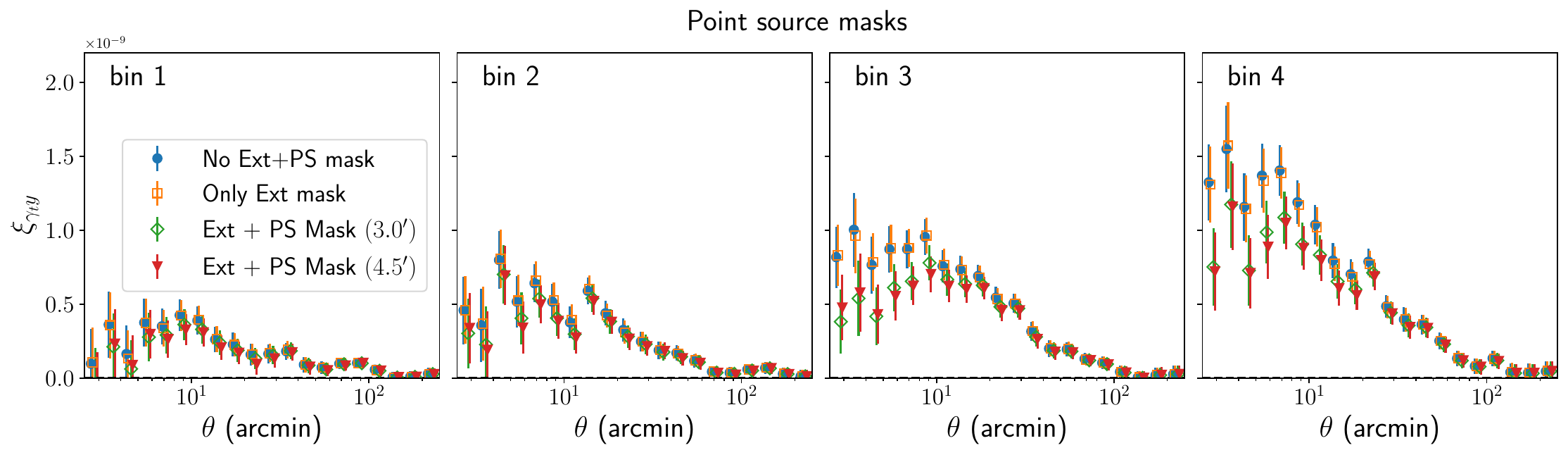}
    }    
    \vfill
    \subfloat[]{%
        \includegraphics[width=0.99\linewidth]{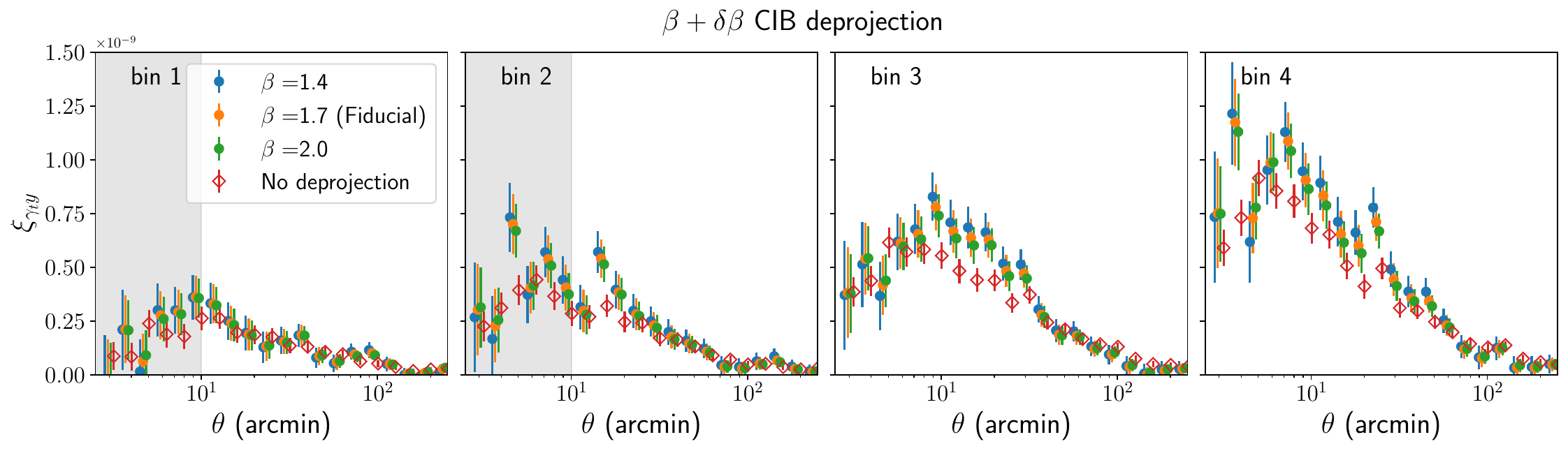}
    }
    \vfill
    \subfloat[]{%
        \includegraphics[width=0.99\linewidth]{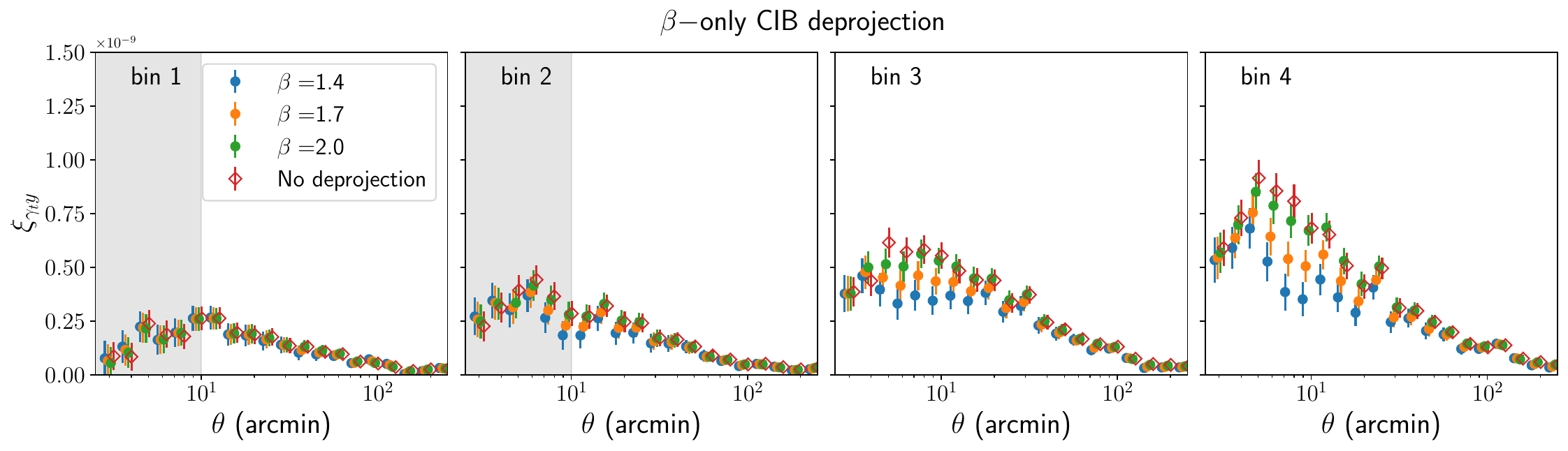}
    }
    
    \caption{Sensitivity of the measurements to the point source treatment and CIB deprojection methodologies. The top row shows the sensitivity of our measurements when using different masks to remove the contribution from point sources as detailed in Appendix~\ref{app:cib}. The middle row shows the first moment deprojection method ($\beta + d\beta$) for different values of $\beta$. The measurement curve for $\beta=1.7$ is our fiducial choice for analysis. The bottom row shows $\beta$-only deprojection, where the measurements vary significantly when changing the value of $\beta$. We also overplot the measurements obtained without any deprojection with significantly smaller error bars, but with clear biases in some of the high-redshift bins.}
    \label{fig:PS_cib_deproj_variations}
\end{figure*}

However, there are significant uncertainties in the effective values of the CIB SED parameters $\beta$ and $T_{\rm CIB}$ that minimize the bias in the cross-correlation measurements of interest due to CIB leakage in the tSZ map. We can see in the bottom panel of Fig.~\ref{fig:PS_cib_deproj_variations} that changing the $\beta$ value has a significant impact on the measurements, especially for the last two bins, which are at higher redshifts and thus more susceptible to CIB contamination.

To make our measurements more robust and less sensitive to the $\beta$ value, we follow the methodology described in \cite{chluba_rethinking_2017, mccarthy_component-separated_2024, coulton_atacama_2024} and additionally deproject a component corresponding to the first moment of the CIB SED relative to $\beta$:
\begin{equation}
    g_{\mathrm{CIB}-d\beta} (\nu) = \frac{A \ln{(\nu/\nu_0)}\left(\frac{\nu}{\nu_0}\right)^{3+\beta}}{\exp \left(\frac{h\nu}{k_B T_{\rm CIB}}\right)-1} \left(\frac{dB(\nu, T)}{dT}\biggr\rvert_{T = T_{\mathrm{CMB}}}\right)^{-1}. 
\end{equation}
As we can see in the middle panel of Fig.~\ref{fig:PS_cib_deproj_variations}, with this $\beta + \delta\beta$ method, our measurements are now robust to the choice of $\beta$ value. Similar conclusions were reached in \citep{mccarthyhill2024} when cross-correlating \emph{Planck}-derived tSZ maps with CMB lensing data, and in  \citep{liu_measurements_2025} when cross-correlating DESI galaxies with the ACT tSZ maps. 

Note that the CIB emission in different frequency channels is not perfectly correlated \citep{lenz_large-scale_2019}. Moreover, the effective integrated emission from low- and high-redshift dusty galaxies can be different compared to the assumed modified blackbody form, which can cause some residual CIB leakage in the tSZ map even with the moment deprojection method. This is an area of active research using both simulations \citep{narayanan_powderday_2021} as well as observations \citep{kampen_atacama_2024}. We defer a detailed analysis of the robustness of the moment deprojection method to these sources of biases to a future study (see also recent work from \citep{surrao2025}). 
\\
\\

\section{shear auto measurements}
\label{app:shear2pt}
\begin{figure*}
\centering
\includegraphics[width=\textwidth]{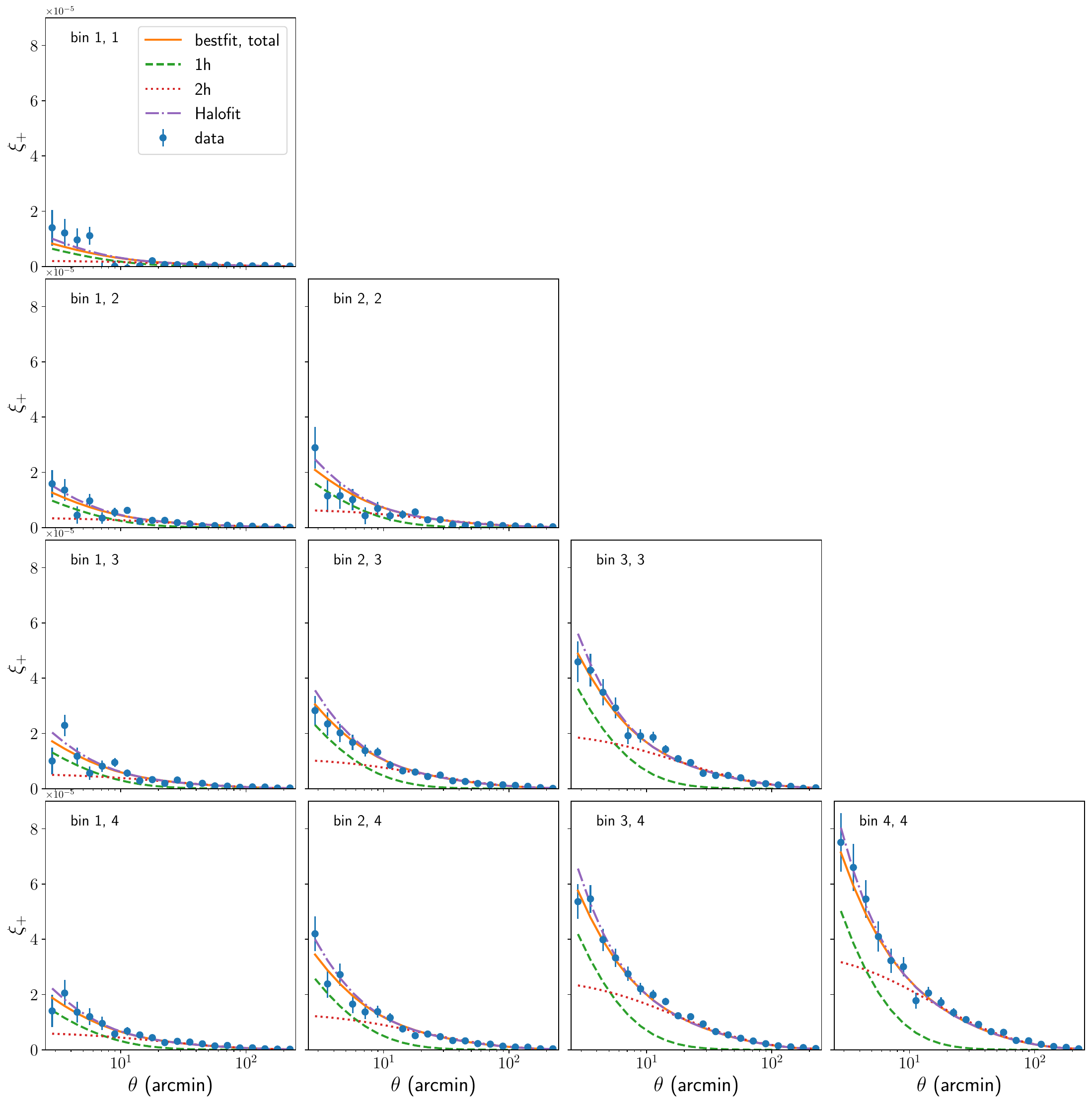}
\caption[]{Measurements and best-fit for shear auto correlation measurement $\xi_{+}$ (see Eq.~\ref{eq:xiplusminus}) for various bin combinations. }
\label{fig:xip_measure}
\end{figure*}

\begin{figure*}
\centering
\includegraphics[width=\textwidth]{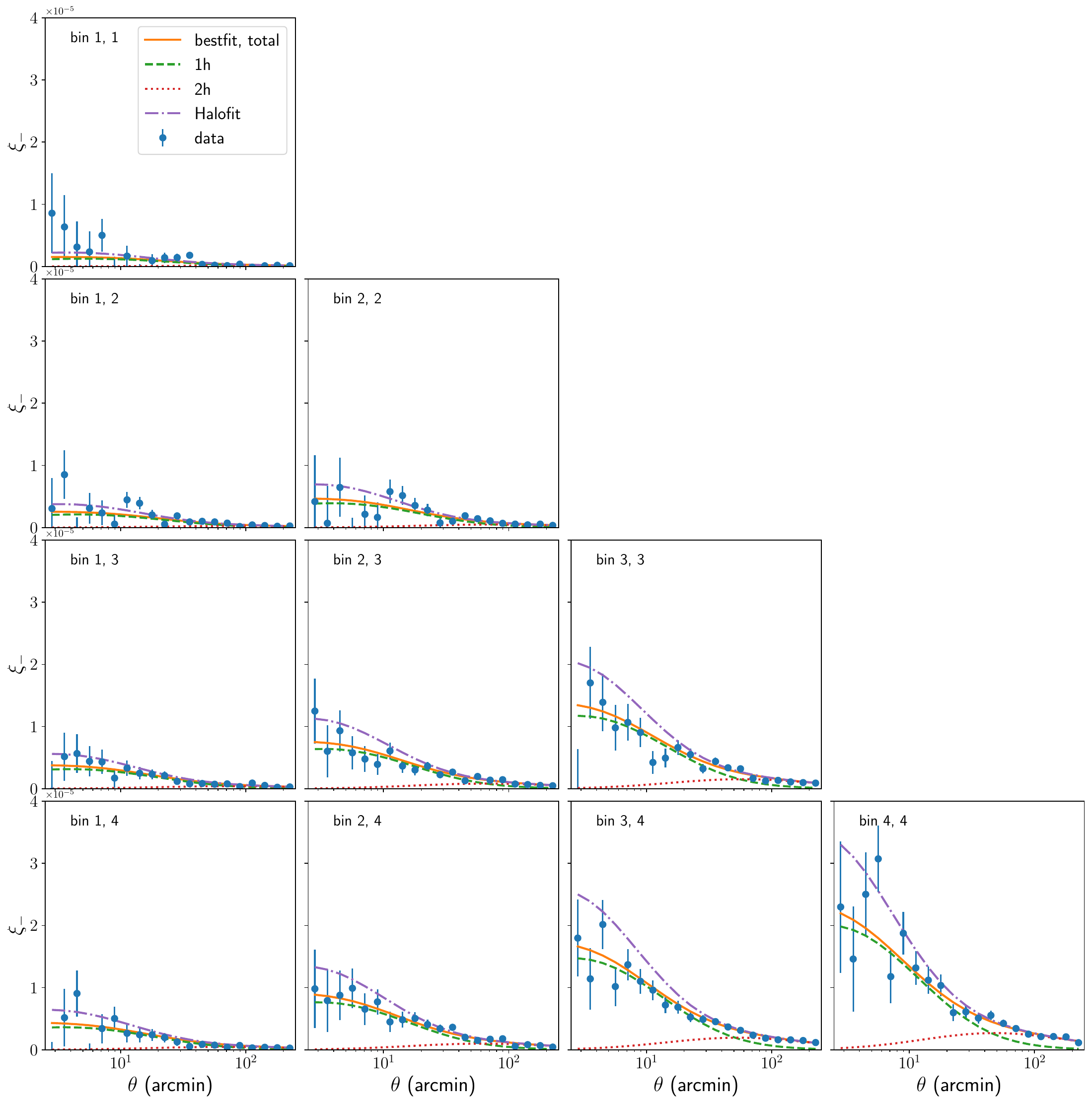}
\caption[]{Measurements and best-fit for shear auto correlation measurement $\xi_{-}$ (see Eq.~\ref{eq:xiplusminus}) for various bin combinations. }
\label{fig:xim_measure}
\end{figure*}

We show the measurements of shear auto correlations, $\xi_{+}$ and $\xi_{-}$ in Fig.~\ref{fig:xip_measure} and Fig.~\ref{fig:xim_measure} respectively. We also show the best-fit curve from our constraints and its split into 1-halo and 2-halo components. We see from the best-fit that $\xi_{+}$ correlations are dominated by the 2-halo contribution, whereas 1-halo term dominates the $\xi_{-}$ correlations. This primarily is caused by $\xi_{+}$ estimator using the zeroth-order Bessel function (see Eq.~\ref{eq:xiplusminus}), which is significantly more localized in its support over the multipole range compared to the fourth-order Bessel function used in the $\xi_{-}$ estimator. Therefore, at a given angular scale, $\xi_{-}$ receives contribution from smaller physical scales, which are dominated by the 1-halo term compared to $\xi_{+}$.

Additionally, we also plot the curve predicted from the \texttt{halofit} fitting function \cite{takahashi_revising_2012}, where cosmology and other parameters are fixed to our best-fit. We see that the small scale measurements of shear auto, particularly for $\xi_{-}$, lie below the \texttt{halofit} prediction. This is in agreement with our findings of increased matter power suppression relative to the predictions from dark matter-only simulations (see \S~\ref{sec:results}) on small scales.

\section{Constraints on all parameters}
\label{app:all_params}

\begin{figure*}
\centering
\includegraphics[width=\textwidth]{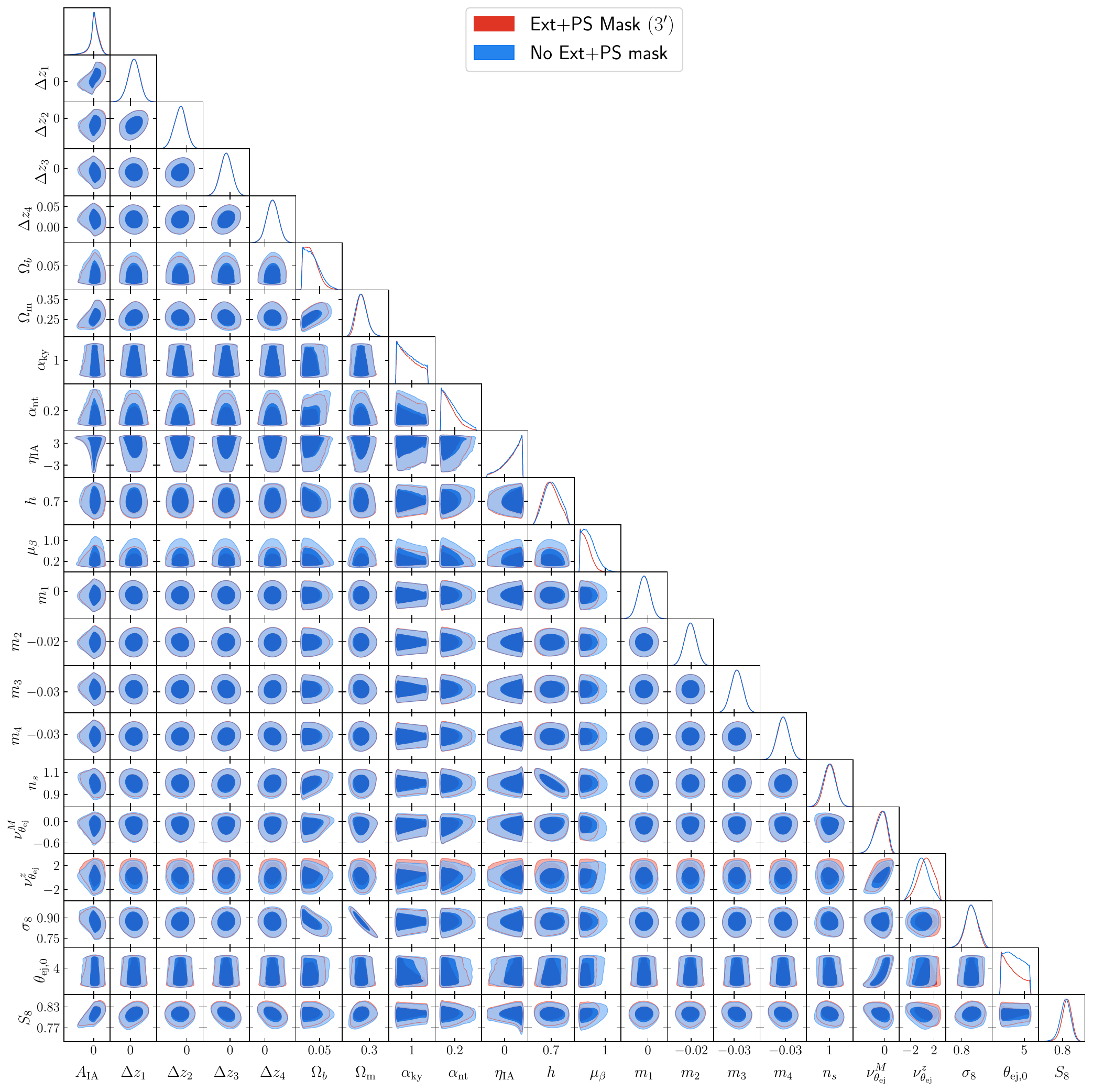}
\caption[]{Constraints on all the parameters varied in this analysis. We show the constraints when masking out the regions around point and extended sources as detected in ACT (red, fiducial constraints) as well as without this mask (blue). }
\label{fig:allparams}
\end{figure*}

All the parameters sampled in this study, along with their priors, were described in Table~\ref{tab:params_all}. In Fig.~\ref{fig:allparams}, we show the constraints on all these parameters with our joint fit to shear auto and shear$\times$tSZ. We show the constraints both when including the mask around point and extended sources (fiducial result) and excluding it and find that our cosmological parameter inferences are robust to this choice. Note that the posteriors on unconstrained parameters span the prior range, which denotes the physical extent of the variation expected in these parameters as described in the main text. We have verified that changing the prior ranges in these unconstrained parameters does not impact our conclusions on the constrained cosmological parameters and astrophysical inferences.

\section{Code validation}
\label{app:code_val}
\begin{figure*}
    \centering
    \includegraphics[width=0.95\textwidth]{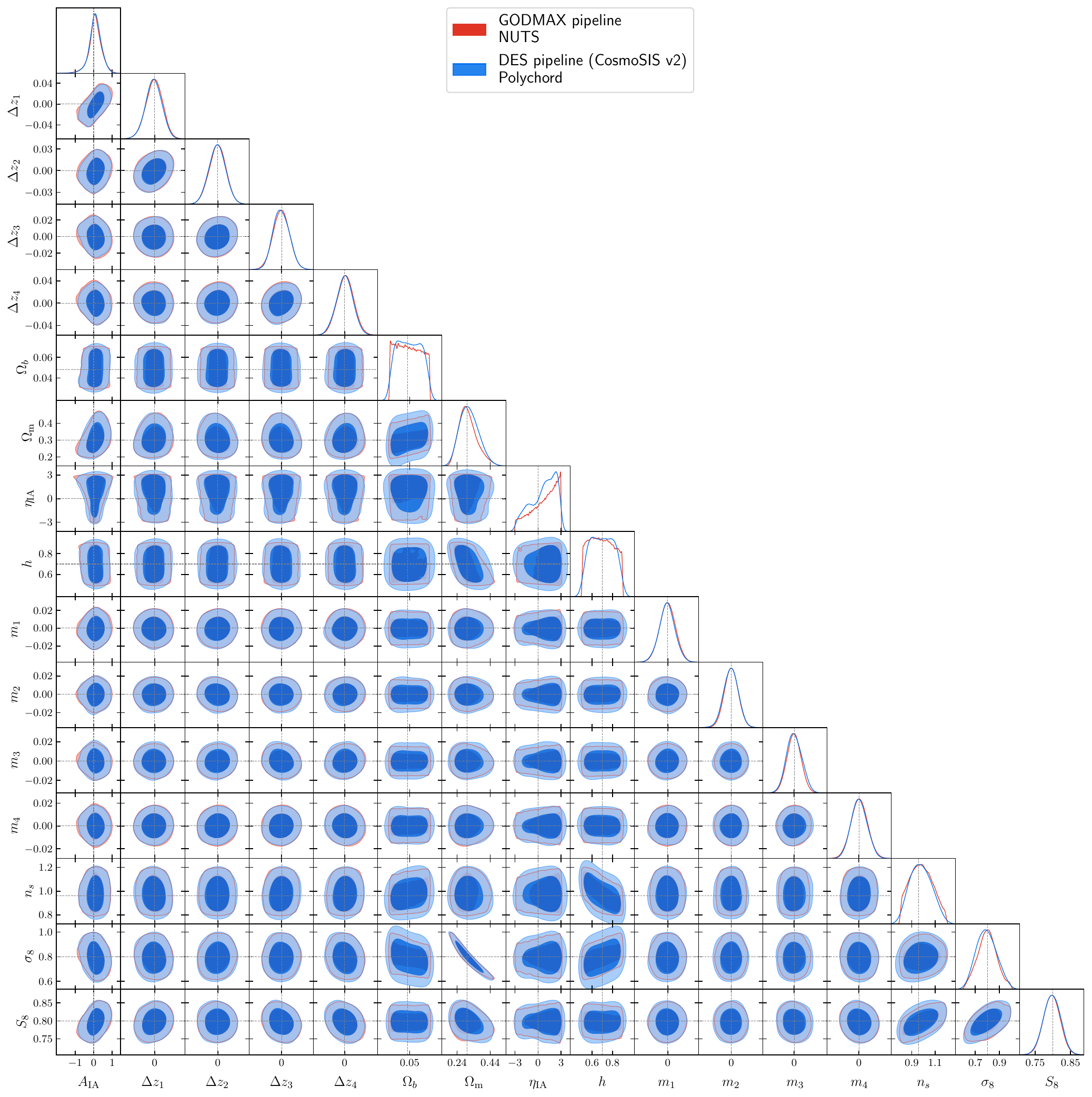}
    \caption[]{Comparison of the parameter constraints from the HMC sampling as used here and Polychord sampling using the cosmosis package when analyzing the same shear auto datavector.}
    \label{fig:code_comp}
\end{figure*}

To validate our sampling scheme and code, we compare the shear auto prediction using the \texttt{halofit} model between our pipeline and the fiducial \texttt{cosmosis} pipeline of DES. We find that the total $\chi^2$ between the two codes for the full datavector is below 1. Moreover, we show the constraints obtained when analyzing a simulated datavector with both the codes in Fig.~\ref{fig:code_comp}, finding consistent constraints. The \texttt{cosmosis} pipeline of DES uses the \cite{handley_polychord_2015} sampler and took 9 hours on 192 cores to converge while the NUTS chain converged in 3 hours on 4 \texttt{Nvidia-H100} GPUs. Note that compared to the model and analysis setup presented in \S~\ref{sec:model}, this is a reduced parameter space and does not require solving the hydrodynamical equations. 

\end{document}